\documentclass{article}

\usepackage[a4paper]{geometry}
\usepackage{graphicx}
\usepackage[colorlinks=true, allcolors=blue]{hyperref}
\usepackage{multirow}
\usepackage{amsmath,amssymb,amsfonts}
\usepackage{amsthm}
\usepackage{mathrsfs}
\usepackage[title]{appendix}
\usepackage{xcolor}
\usepackage{textcomp}
\usepackage{manyfoot}
\usepackage{booktabs}
\usepackage{algorithm}
\usepackage{algorithmicx}
\usepackage{algpseudocode}
\usepackage[most]{tcolorbox}
\tcbuselibrary{skins,breakable,listings,breakable}
\usepackage{markdown}

\usepackage{listings}
\makeatletter
\def\lst@makecaption{
  \def\@captype{table}%
  \@makecaption
}
\makeatother

\usepackage[T1]{fontenc}

\usepackage{cleveref}
\usepackage{comment}
\usepackage{colortbl}
\usepackage{dsfont}
\usepackage{tikz}
\usetikzlibrary{positioning,decorations.pathreplacing,fit,calc,arrows,tikzmark}
\usepackage{bigstrut}
\usepackage{caption}
\usepackage{subcaption}
\usepackage{tabularx,ragged2e}
\usepackage{longtable,booktabs,threeparttablex}
\usepackage{xltabular}
\usepackage{xtab,afterpage}
\usepackage{orcidlink}

\usepackage[
    style=authoryear,
    natbib=true
]{biblatex}
\usepackage{software-biblatex}
\title{Favia: Forensic Agent for Vulnerability-fix Identification and Analysis}

\author{
  Andr\'e Storhaug\thanks{This work was partially done during research visit at CSIRO Data61.} \orcidlink{0000-0002-5321-7196}\\
  Norwegian University of Science and Technology, Trondheim, Norway\\
  \texttt{andre.storhaug@ntnu.no}
  \and
  Jiamou Sun \orcidlink{0000-0002-5212-7068}\\
  CSIRO's Data61, Canberra, ACT, Australia\\
  \texttt{Frank.Sun@data61.csiro.au}
  \and
  Jingyue Li \orcidlink{0000-0002-7958-391X}\\
  Norwegian University of Science and Technology, Trondheim, Norway \\
  \texttt{jingyue.li@ntnu.no}
}

\date{}

\begin{document}
\maketitle

\begin{abstract}
Identifying vulnerability-fixing commits corresponding to disclosed CVEs is essential for secure software maintenance but remains challenging at scale, as large repositories contain millions of commits of which only a small fraction address security issues. Existing automated approaches, including traditional machine learning techniques and recent large language model (LLM)-based methods, often suffer from poor precision-recall trade-offs. Frequently evaluated on randomly sampled commits, we uncover that they are substantially underestimating real-world difficulty, where candidate commits are already security-relevant and highly similar. We propose Favia, a forensic, agent-based framework for vulnerability-fix identification that combines scalable candidate ranking with deep and iterative semantic reasoning. Favia first employs an efficient ranking stage to narrow the search space of commits. Each commit is then rigorously evaluated using a ReAct-based LLM agent. By providing the agent with a pre-commit repository as environment, along with specialized tools, the agent tries to localize vulnerable components, navigates the codebase, and establishes causal alignment between code changes and vulnerability root causes. This evidence-driven process enables robust identification of indirect, multi-file, and non-trivial fixes that elude single-pass or similarity-based methods. We evaluate Favia on CVEVC, a large-scale dataset we made that comprises over 8 million commits from 3,708 real-world repositories, and show that it consistently outperforms state-of-the-art traditional and LLM-based baselines under realistic candidate selection, achieving the strongest precision-recall trade-offs and highest F1-scores.
\end{abstract}

\section{Introduction}
\label{sec:introduction}

Security vulnerabilities in open-source software pose a persistent and growing threat to modern software ecosystems. Once a vulnerability is disclosed—typically through a Common Vulnerabilities and Exposures (CVE) entry, developers, security teams, and downstream users must quickly identify the corresponding patch commit in the affected repository to assess exposure, apply fixes, and propagate updates. However, locating vulnerability-fixing commits remains a challenging and time-consuming task. Large repositories such as the Linux kernel contain millions of commits, of which only a tiny fraction correspond to security patches, making manual inspection or exhaustive analysis infeasible.

Prior research has explored automated vulnerability fix detection using a range of techniques, including handcrafted rules \citep{sliwerski2005when}, traditional machine learning \citep{wang2019detecting}, and deep learning models \citep{hoang2019patchnet:}. While these approaches have demonstrated promising results, they suffer from poor precision-recall trade-offs. Conservative models miss most true patches, while aggressive models generate large numbers of false positives.

Recent advances in large language models (LLMs) have enabled more semantically rich reasoning over code and natural language \citep{yang2024swe-agent}, leading to LLM-based frameworks for vulnerability fix detection \citep{yang2025code, wu2025commitshield}. These systems leverage pretrained knowledge and structured prompting to reason about CVE descriptions and code changes. However, existing LLM-based approaches typically operate in a single-pass manner over shallow context, making them prone to superficial associations such as keyword overlap or file-name similarity. Further, most evaluations rely on simplified or randomly sampled candidate sets which substantially underestimate the difficulty of real-world deployment scenarios. As most commits in a repository are not security-relevant and not very similar to the true patch, it is easy to distinguish between a patch and non-patch. As a result, reported performance often fails to reflect behavior under realistic operating conditions.

In this work, we propose \textbf{Favia}, a \emph{forensic agent} for vulnerability-fix identification and analysis. Favia adopts a hybrid design that combines scalable candidate ranking with deep, agent-based semantic reasoning. We adopt this hybrid design to balance scalability and analytical depth. Lightweight candidate ranking efficiently narrows the vast search space of commits to a manageable subset, while agent-based semantic reasoning enables fine-grained, evidence-driven analysis of each candidate. Given a CVE and a ranked list of candidate commits, Favia instantiates an LLM-based agent within a code environment reflecting the repository state prior to each commit. Using a iterative reasoning loop, the agent retrieves CVE and CWE reports, localizes affected components, navigates the codebase, and explicitly correlates code changes with the vulnerability’s root cause, entry points, and impact. This iterative, evidence-driven process enables Favia to reason about indirect and non-trivial fixes that are difficult to capture with similarity-based or single-pass methods.

To rigorously evaluate Favia, we construct \textbf{CVEVC}, a large-scale dataset comprising over 8 million commits drawn from 3,708 real-world repositories associated with publicly disclosed CVEs. Using this dataset, we compare Favia against state-of-the-art traditional and LLM-based baselines under two evaluation settings: (i) a \emph{random} setting commonly used in prior work, and (ii) a \emph{realistic} setting where all repository commits are candidates. In addition to standard precision, recall, and F1-score metrics, we perform an in-depth analysis of agent behavior, tool usage, failure modes, and computational cost.

Our results show that Favia consistently outperforms state-of-the-art traditional and LLM-based baselines. On the realistic dataset, Favia achieves the highest F1-scores across all tested models by preserving near-perfect recall (up to 0.98) while substantially improving precision over prior LLM-based approaches by up to 92\%. This is achieved by perform iterative, evidence-grounded code navigation, enabling reliable causal alignment between CVE root causes and commit-level changes. We also demonstrate how randomly sampled commits dramatically inflate performance for all methods, with F1-scores increasing by up to 95\%, masking real-world difficulty and narrowing apparent gaps between approaches. Failure analysis shows that over 85\% of agent errors stem from superficial association between code changes and CVE, or misinterpretation of CVE root causes rather than limited exploration or tool access.

Our contributions are threefold:
\begin{itemize}
    \item We introduce Favia, an agent-based framework that performs iterative, evidence-grounded reasoning for vulnerability-fix identification.
    \item We provide a high-quality dataset designed to support realistic evaluation of vulnerability-fix detection.
    \item We present a comprehensive empirical study showing that agent-based reasoning substantially improves precision–recall trade-offs under realistic conditions, while also revealing key failure modes and efficiency trade-offs.
\end{itemize}

The rest of the paper is organized as follows. \Cref{chap:preliminaries} explains required background theory. \Cref{chap:related-work} reviews related work. \Cref{chap:approach} details our agentic approach for patch classification. \Cref{chap:related-work} introduces related work. We explain our experimental design in \Cref{chap:experimental-design}. \Cref{chap:experimental-results} presents the experimental results.  \Cref{chap:discussion} discusses our results and limitations. \Cref{chap:conclusion} concludes the study and proposes future work.

\section{Preliminaries}
\label{chap:preliminaries}

\subsection{Patch Commit}
\label{section:patch-commit}
A patch commit refers to a specific change in a software repository that addresses and resolves a known vulnerability. These commits are typically associated with security advisories, such as CVEs (Common Vulnerabilities and Exposures), and are intended to mitigate or eliminate exploitable flaws in the codebase. Patch commits may involve direct modifications to the vulnerable code or indirect changes that affect related components, configurations, or control flows. As modern software is composed of deeply nested dependencies, when an upstream vulnerability is addressed, downstream maintainers must precisely identify the corresponding patch commits to assess exposure, backport fixes, and release secure updates \citep{li2017security}. Thus, identifying patch commits is a critical task in software security, as it enables downstream systems to track, verify, and propagate security fixes across dependent projects and distributions. Failing to do so has led to billion-dollar losses from malicious sabotage, such as the infamous Log4Shell \citep{doll2025unraveling} and SolarWinds \citep{Yang2022} attacks.

\subsection{Vulnerability Fix Detection}
\label{section:vulnerability-fix-detection}

The task of vulnerability fix detection, also referred to as patch detection, is the process of identifying which commits in a version-controlled repository correspond to security patches. This problem is motivated by the need to automate vulnerability tracking and remediation in large-scale software ecosystems. A key challenge is that all commits in a repository are potential candidates, which makes the search space prohibitively large. For instance, as of January 2006, the Linux kernel contains over 1.4 million commits. Only a small fraction of which are related to security fixes. As of December 30th, 2025, 9981 CVEs are reported in the linux-cve mailing list \citep{linux_cve_announce}.

Moreover, vulnerability fixes are often not trivial. While some vulnerabilities, such as integer overflows, may be resolved with localized checks or type constraints, others, such as logic flaws, race conditions, or privilege escalation bugs, require broader contextual reasoning. These fixes may span multiple files, involve indirect control paths, or rely on subtle behavioral changes that are not immediately evident in the diff. As a result, distinguishing true vulnerability patches from unrelated or cosmetic changes demands a deep understanding of both the vulnerability and the surrounding codebase.

\begin{lstlisting}[
    language=diff,
    style=diff,
    caption={Patch commit 67f2cdd\protect\footnotemark for CVE-2014-100019.},
    label=lst:patch-CVE-2014-100019,
    numbers=none,
    frame=none,
    breaklines=true,
    belowskip=0pt,
    escapechar=§]
diff --git a/Pomm/Converter/PgLTree.php b/Pomm/Converter/PgLTree.php
index 6377f8e..e408c7b 100644
--- a/Pomm/Converter/PgLTree.php
+++ b/Pomm/Converter/PgLTree.php
@@ -27,6 +27,9 @@ class PgLTree implements ConverterInterface
\end{lstlisting}
\begin{lstlisting}[
    language=diff,
    style=diff,
    aboveskip=0pt,
    frame=bottomline,
    breaklines=true,
    firstnumber=27,
    escapechar=§]
      */
       public function toPg($data, $type = null)
       {
 -         return sprintf("'%s'::ltree", join('.', $data));
 +         $data = join('.', $data);
 +         $data = str_replace("'", "''", $data);
 +         $data = str_replace("\\", "\\\\", $data);
 +         return sprintf("'%s'::ltree", $data);
       }
   }
\end{lstlisting}

\footnotetext{\url{https://github.com/chanmix51/Pomm/commit/67f2cddd1cd79153ebb37228472d8962a541a6fd}}

To illustrate this challenge, consider a SQL injection vulnerability in the PgLTree converter of POMM (\href{https://nvd.nist.gov/vuln/detail/CVE-2014-100019}{CVE-2014-100019}), which allows remote attackers to execute arbitrary SQL commands due to improper handling of user-controlled input. The corresponding patch (shown in \cref{lst:patch-CVE-2014-100019}) mitigates the vulnerability by introducing proper string escaping (lines 31-34), directly addressing the root cause of the SQL injection. Such fixes are relatively easy to detect, as the mitigation mechanism clearly aligns with the vulnerability description. However, in practice, a vulnerability-fixing commit may closely resemble a non-patching commit, making the two difficult to distinguish. For example, \cref{lst:non-patch-CVE-2014-100019} shows another commit from the same repository that appears to address a similar SQL-related issue but modifies a different, albeit superficially plausible, code location. Specifically, line 75 is changed into line 76, wrapping the \texttt{array\_shift(\$elts)} in a function call to \texttt{stripcslashes}. Despite its similarity (escaping slashes), this commit does not resolve the reported vulnerability. Correctly avoiding a false positive in this case requires reasoning about the presence, or absence, of \texttt{PgLTree}-related logic. In particular, a human would typically search for references to \texttt{PgLTree} and assess whether the changes meaningfully interact with the vulnerable \texttt{PgLTree} conversion path. If no such relationship exists, the commit is unlikely to constitute the true vulnerability fix, even if the change appears relevant at a surface level.

\begin{lstlisting}[
    language=diff,
    style=diff,
    caption={Commit d84dff2\protect\footnotemark from chanmix51/Pomm.},
    label=lst:non-patch-CVE-2014-100019,
    numbers=none,
    frame=none,
    breaklines=true,
    belowskip=0pt,
    escapechar=§]
diff --git a/Pomm/Converter/PgEntity.php b/Pomm/Converter/PgEntity.php
index 15eb808..f83aaf6 100644
--- a/Pomm/Converter/PgEntity.php
+++ b/Pomm/Converter/PgEntity.php
@@ -70,7 +70,7 @@ class PgEntity implements ConverterInterface
\end{lstlisting}
\begin{lstlisting}[
    language=diff,
    style=diff,
    aboveskip=0pt,
    frame=bottomline,
    breaklines=true,
    firstnumber=70,
    escapechar=§]
         throw new NullPointerException();
       }
           $fields = array();
           foreach ($this->map->getFieldDefinitions() as $field_name => $pg_type)
           {
 -             $fields[$field_name] = array_shift($elts);
 +             $fields[$field_name] = stripcslashes(array_shift($elts));
           }
   
           if (count($elts) > 0)
\end{lstlisting}

\footnotetext{\url{https://github.com/chanmix51/Pomm/commit/d84dff2165d5900296cf47d8923ff6333f5638a2}}

\section{Related Work}
\label{chap:related-work}

\subsection{Vulnerability Fix Detection}
\label{sec:related-work-vfd}
Vulnerability fix detection has been a prominent research area, encompassing static and dynamic analysis techniques as well as machine learning approaches. Early studies primarily relied on rule-based systems and manually crafted features to identify security-relevant commits \citep{mockus2000identifying, sliwerski2005when}. With the growth of large-scale software projects, classical machine learning and natural language processing techniques were increasingly applied to commit messages, code diffs, and issue-tracking artifacts \citep{wang2019detecting}. More recent work has leveraged deep learning \citep{hoang2019patchnet:}, graph-based representations \citep{wu2022-enhancing}, and large language models (LLMs) to capture semantic, syntactic, and structural characteristics of code changes \citep{yang2025code}. In the following sections, we provide a structured overview of this evolution.

\subsubsection{Heuristic and Metadata-Based Detection}

Early works primarily relied on heuristic signals derived from commit metadata and simple code change characteristics in order to identify patches. Several studies searched for keywords like ``fix'' and ``bug'' in log messages to find vulnerability patches \citep{mockus2000identifying, sliwerski2005when, dallmeier2007extraction, kim2008predicting}. \citet{tian2012identifying} extended the analysis to use both commit messages and code changes. These approaches extracted manually designed ``facts'' from text and code, such as the number of added loops or function calls, to build discriminative models. This line of work naturally evolved toward vulnerability identification by incorporating repository-level metadata, including committer experience, contribution history, and submission timing, which were used to train Support Vector Machine (SVM) classifiers to support security auditing workflows \citep{perl2015vccfinder:}. These methods are brittle to vocabulary changes and developer practices, and their reliance on shallow, manually engineered signals limits generalization and recall for non-explicit or silent vulnerability fixes.

\subsubsection{Classical and Semi-Supervised Learning Approaches}

Subsequent research expanded beyond kernel-specific heuristics to broader software ecosystems, leveraging classical machine learning and natural language processing to analyze developer-written artifacts. Studies utilized commit messages and bug reports collected from issue-tracking platforms such as JIRA and Bugzilla to detect security-relevant changes \citep{zhou2017automated}. To reduce analyst burden in industrial settings, later systems combined independent classifiers over log messages and patches, prioritizing high precision to limit false positives \citep{sabetta2018-a-practical}. In parallel, researchers began addressing the challenge of ``secret'' security patches-fixes applied without public disclosure-by applying ensemble-based classifiers to capture syntactic and semantic indicators of vulnerability fixes in code changes \citep{wang2019detecting}. While more flexible than heuristics, these approaches depend heavily on labeled data and surface-level textual cues, making them less effective for semantically complex fixes and previously unseen vulnerability patterns.

\subsubsection{Deep Learning for Commit and Code Change Modeling}

The adoption of deep learning marked a turning point in vulnerability-fixing commit detection. Hierarchical neural architectures such as PatchNet were proposed to learn semantic representations that reflect the sequential and hierarchical structure of code diffs \citep{hoang2019patchnet:}. Around the same time, the notion of ``safe patches'' emerged, using symbolic execution and interpretation to identify fixes that constrain input spaces without altering intended program behavior, enabling faster downstream adoption \citep{machiry2020spider:}. Further advances incorporated representation learning guided by commit messages, as exemplified by CC2Vec, which employs attention mechanisms to jointly model code changes and their textual descriptions while capturing structural dependencies \citep{hoang2020cc2vec:}. Despite improved representation learning, deep neural models often require large, curated training datasets and struggle with interpretability and transferability across projects and programming languages.

\subsubsection{Structural and Graph-Based Representations}

To better capture program structure, later work emphasized syntax-aware and graph-based representations of code changes. Approaches such as Commit2Vec \citep{cabrera-lozoya2021commit2vec:} modeled the symmetric difference between Abstract Syntax Trees (ASTs) before and after a commit to encode fine-grained structural modifications. RNN-based models like PatchRNN \citep{wang2021patchrnn:} extracted syntax-level features from both diffs and commit messages to improve detection performance. Similarly, the SPI \citep{zhou2021spi:} system employed composite neural architectures to learn distinctions at the statement level within code revisions. More recently, graph neural networks have been applied to enriched representations such as Patch Code Property Graphs (PatchCPGs), enabling systems like GraphSPD \citep{wang2023graphspd:} and E-SPI \citep{wu2022-enhancing} to capture semantic relationships and control/data-flow dependencies. Other tools, including VFFINDER \citep{nguyen2023vffinder:}, rely on annotated ASTs to represent structural changes with high precision, while transformer-based models fine-tuning CodeBERT have been used to detect silent vulnerability fixes across projects and programming languages, e.g. the popularly used VulFixMiner \citep{zhou2021finding}. Although structurally precise, graph- and syntax-based methods incur substantial preprocessing and computational overhead, which hampers scalability and adoption in large, rapidly evolving codebases.

\subsubsection{Multi-View, Traceability, and Ranking-Based Methods}

A persistent challenge in the literature is the lack of explicit links between commits and external vulnerability or issue-tracking data. To address this, systems, such as HERMES by \citet{nguyen-truong2022hermes:}, introduced link recovery techniques to enrich commit context and improve vulnerability-fix detection. In parallel, ranking-based methods reframed the problem as correlating vulnerabilities with commits based on identifiers, file locations, and textual similarity. Approaches like PatchScout and VCMATCH rank candidate patches by measuring alignment between vulnerability descriptions and code changes \citep{tan2021locating,wang2022vcmatch:}. To handle tangled commits, multi-granularity frameworks such as MiDas analyze changes at multiple levels-commit, file, hunk, and line-using dedicated neural components \citep{nguyen2023multi-granularity}. Efforts to improve usability and interpretability include VulCurator, which produces ranked lists of likely fixes using BERT-based representations of text and code \citep{nguyen2022vulcurator:}, and methods that generate aspect-level explanations such as root causes and impacts alongside predictions \citep{sun2023silent}. These techniques rely on the availability and quality of auxiliary artifacts and recovered links, and their performance degrades when vulnerability descriptions are incomplete, ambiguous, or weakly aligned with code changes.

\subsubsection{Large Language Model–Based Approaches}

The recent advances in Large Language Models (LLMs) has enabled reasoning about vulnerability fixes at a higher semantic level. Two-phase frameworks such as PatchFinder combine hybrid retrieval mechanisms for coarse filtering with semantic re-ranking to identify relevant patches efficiently \citep{li2024patchfinder:}, reporting recall@10 of 80.63\%. Frameworks such as CompVPD \citep{chen2024compvpd:}, LLM4VFD \citep{yang2025code} and CommitShield \citep{wu2025commitshield} leverage pretrained LLMs to reason over code changes in natural language. CompVPD introduces iterative refinement through precise context selection and human validation signals to address generalization limitations \citep{chen2024compvpd:}. LLM4VFD employs pretrained language models to align vulnerability descriptions with code changes, enabling more semantically informed identification of vulnerability-fixing commits \citep{yang2025code}. CommitShield extends this paradigm by incorporating Chain-of-Thought reasoning and program slicing to provide structured contextual evidence for its predictions, allowing for iterative refinement and improved precision \citep{wu2025commitshield}.

Although LLM-based approaches show significant improvements over traditional learning-based methods, they introduce challenges in complexity, computational cost, and scalability. Running large pretrained models with reasoning-intensive pipelines on repositories containing millions of commits, such as the Linux kernel, remains expensive. Consequently, evaluations are often limited to curated or constrained experimental settings, leaving open questions about robustness and real-world effectiveness. In contrast, our approach integrates the efficiency and scalability of classical machine learning techniques with the adaptive reasoning capabilities of an LLM-based agent system, providing human-like contextual understanding while remaining practical for large-scale software ecosystems.

\section{Favia Approach}
\label{chap:approach}

Our vulnerability fix detection framework Favia identifies vulnerability-fixing commits through a two-stage process. It first applies a fast, coarse-grained ranking over all commits in an affected repository to surface a small set of candidate patches. Each candidate is then examined in depth by an Large Language Model (LLM) agent that performs detailed semantic reasoning over the codebase and vulnerability context to determine whether the commit constitutes a true vulnerability fix. This hybrid design balances scalability with semantic depth, enabling efficient filtering of potential patch commits followed by rigorous contextual analysis.

\subsection{Candidate Ranking}

\begin{figure}[htbp]
    \centering
    \includegraphics[scale=0.75]{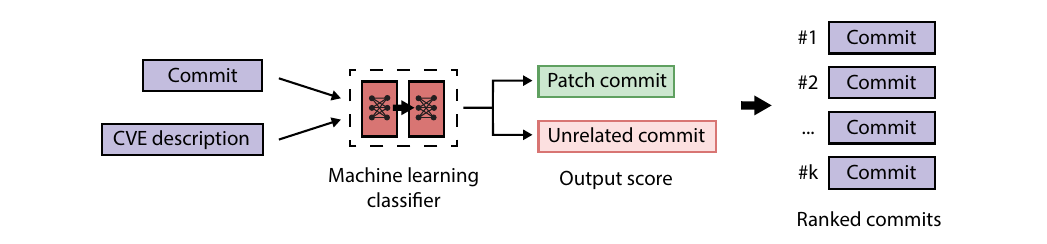}
    \caption{Top-n ranking using machine learning classifier.}
    \label{fig:candidate-ranking}
\end{figure}

The first stage aims to drastically reduce the search space of potential patch commits. Given that repositories such as the Linux kernel contain over 1.4 million commits, exhaustive evaluation with heavy Large Language Models (LLMs) is computationally infeasible. We therefore employ a more conventional machine learning classifier to rank commits based on their likelihood of being vulnerability fixes. These classifiers may also include static and dynamic analysis tools, as summarized in \cref{chap:related-work}. Specifically, we chose to use PatchFinder by \citet{li2024patchfinder:} (\cref{sec:patchfinder}) as our classifier for commit candidate ranking because it is the leading approach among efficiency-oriented methods for vulnerability-fix identification. However, any sufficiently efficient classifier may be used. The only requirement is that the classifier selects the top-k candidate commits for further inspection. 

\Cref{fig:candidate-ranking} illustrates the first stage of our Favia approach. Using the CVE description, along with a commit's message and code changes, these are classified with a machine learning classifier, outputting a classification score of the commit being the true patching commit for the CVE. This is repeated for every commit in the repository. All repository commits are then ranked according to score. Finally, only the top-k candidates are kept for the next stage - agent-based classification. The value of k can be tuned based on resource constraints and desired recall. It is important to note that if the true patch is not included in the top-k, the subsequent evaluation stage cannot recover it.

\subsection{Agent-based Classification}
\label{chap:agent-based-evaluation}

\begin{figure}[htbp]
    \centering
    \includegraphics[scale=0.75]{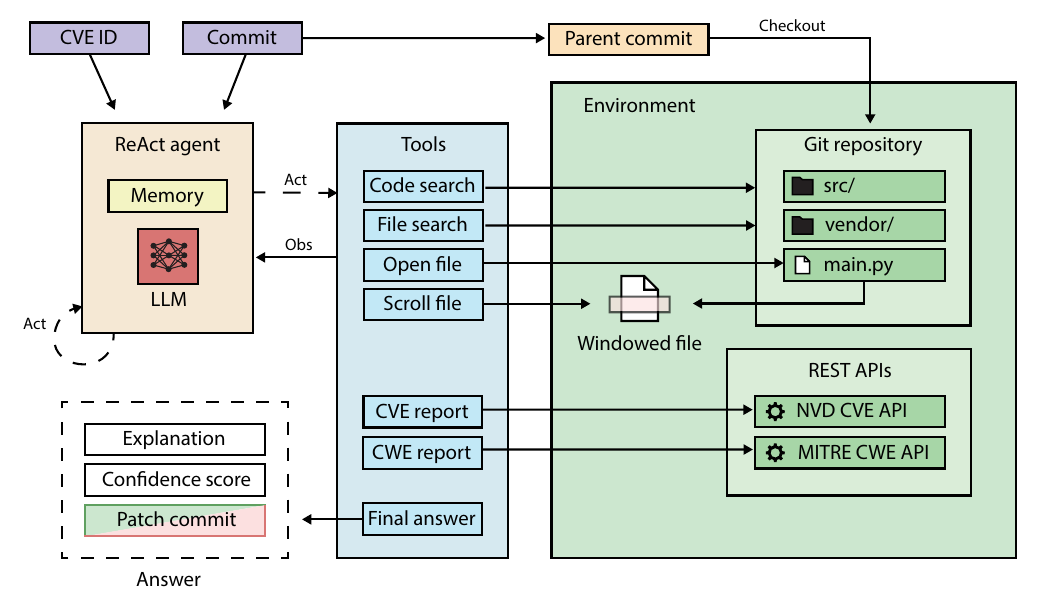}
    \caption{Agent classification for each top-n ranked commits.}
    \label{fig:agent-based-classification}
\end{figure}

\Cref{fig:agent-based-classification} illustrates the second stage of our Favia approach. This stage involves a reasoning agent tasked with evaluating each of the top-k candidates. To facilitate this process, we adopt the ReAct \citep{yao2023reactsynergizingreasoningacting} framework, which integrates reasoning and acting in a unified agent architecture. ReAct enables the agent to iteratively decompose complex evaluation tasks into subgoals, alternating between chain-of-thought reasoning and tool-based interactions. This allows the agent to simulate human-like analysis, retrieve relevant context, and dynamically refine its understanding. Specifically, we implement the agent using the smolagents \citep{smolagents:1.21.2} library from \citet{huggingface}. We keep the standard system prompt that allows for generic tool usage as default (see \cref{fig:system-template}).

Each candidate commit is presented to the agent with the goal of determining whether it constitutes a valid vulnerability patch for a given CVE. We structure this task into three segments:
\begin{enumerate}
    \item \textbf{Understand the Vulnerability}: The agent begins by analyzing the associated CVE and CWE reports to comprehend the nature of the vulnerability.
    \item \textbf{Analyze the Commit Changes}: The agent inspects the diff to identify what modifications were introduced.
    \item \textbf{Correlate Code Changes with the CVE}: The agent evaluates whether the changes effectively mitigate the described vulnerability.
\end{enumerate}
We formulate these segments into the task prompt available in \cref{fig:task-template}. The expected output of the system is threefold. First, the agent produces an explanation of its reasoning, which encourages the agent to consolidate and justify its answer and allows for straightforward qualitative evaluation. Second, the agent reports a confidence score using a 5-point ordinal confidence scale, where 1 indicates no confidence and 5 indicates full confidence. Third, the agent produces a final verdict, ``true'' or ``false'', indicating whether the commit constitutes the true vulnerability patch.

As described in \cref{section:vulnerability-fix-detection}, vulnerability fixes are hard to distinguish from unrelated or cosmetic changes. It is therefore necessary for the agent to be able to inspect the underlying codebase. To support this, we instantiate the agent within a code environment (described in \cref{sec:environment} that reflects the repository state prior to the commit. This enables the agent to locate the vulnerable logic and determine whether the proposed changes address it. The agent is equipped with a suite of tools, detailed in \cref{sec:tools}. These tools allow the agent to navigate the codebase, retrieve relevant context, and perform targeted inspections. By combining structured reasoning with interactive tool use, the agent achieves a level of semantic depth that surpasses conventional classifiers. We allow the agent to run a maximum of 20 steps. If the agent has not finished in 20 steps, the agent is exempted to produce the final answer.

\subsubsection{Environment}
\label{sec:environment}
To enable deep semantic evaluation of candidate commits, we instantiate the agent within a simulated code environment. This environment reflects the state of the repository immediately prior to the commit under investigation. By providing access to the pre-change codebase, the agent can inspect the original logic, identify potential vulnerabilities, and assess whether the proposed diff mitigates the issue. This setup is particularly important for evaluating indirect fixes, where the vulnerability may reside in a transitive dependency or auxiliary function not explicitly modified in the commit.

As seen from \cref{fig:agent-based-classification}, the environment is designed to support interactive exploration, allowing the agent to reason about the codebase in a manner similar to a human analyst. It is equipped with a suite of tools that facilitate targeted inspection, contextual retrieval, and structured reasoning. These tools are exposed as Python functions. Through code generation, the agent can generate function calls to execute any number of tools dynamically on-demand as it sees fit, based on the identification of missing information. For example, the agent may invoke a file search via
\texttt{file\_search(query="FileOutputStream.java")}.

\subsubsection{Tools}
\label{sec:tools}
In the following paragraphs we describe the set of tools available to the agent during vulnerability fix detection analysis. Each tool provides access to a specific type of information, ranging from structured vulnerability metadata to repository-level code inspection, and is explicitly invoked by the agent as part of its reasoning process.

\paragraph{CVE report}
\label{sec:tools-cve-report}
The CVE Report tool retrieves structured vulnerability metadata for a given CVE (Common Vulnerabilities and Exposures) identifier. The information is returned in Markdown format and includes the following key sections:
\begin{itemize}
    \item \textbf{CVE Details}: Includes the CVE ID, source identifier, publication and last modification dates, and current vulnerability status.
    \item \textbf{Known Exploited Status}: Indicates whether the vulnerability is known to be actively exploited in the wild.
    \item \textbf{Scores}: Severity metrics such as CVSS (Common Vulnerability Scoring System) scores, which quantify the impact and exploitability of the vulnerability.
    \item \textbf{Description}: A natural language summary of the vulnerability, filtered by language preference (e.g., English).
    \item \textbf{Weaknesses}: Associated CWE (Common Weakness Enumeration) entries that categorize the vulnerability type.
    \item \textbf{Configurations}: Affected software configurations, platforms, and version constraints.
\end{itemize}
This tool enables the agent to understand the nature and scope of the vulnerability, identify related weaknesses, and contextualize the commit under evaluation. Because the output is rendered in Markdown, it is easily readable and structured for parsing or display. Additionally, this tool serves as a diagnostic mechanism: if the agent bypasses this lookup and prematurely concludes patch validity, it may indicate memorization or data leakage from pretraining. By requiring explicit invocation of this tool, we ensure that the agent’s reasoning is grounded in retrieved evidence rather than latent knowledge.

\paragraph{CWE report}
The CWE Report tool retrieves structured information about a CWE (Common Weakness Enumeration) entry and returns it in Markdown format. The agent uses this tool to understand the general class of vulnerability associated with a given CVE, such as buffer overflows, improper input validation, or race conditions. This contextual knowledge helps the agent reason about the nature of the fix and whether the code changes align with the expected mitigation strategy. It could also find other CVEs with same vulnerability, and consecutively look them up using the CVE report tool.
The CWE report includes the following key sections:
\begin{itemize}
    \item \textbf{Description and Extended Description}: A concise summary of the weakness and its broader implications.
    \item \textbf{Common Consequences}: Typical impacts of the weakness, categorized by scope (e.g., confidentiality, integrity) and effect (e.g., data corruption, privilege escalation).
    \item \textbf{Relationships}: Links to related CWEs, including parent, child, or peer relationships.
    \item \textbf{Content History}: Metadata about the evolution of the CWE entry, including submission and modification records.
\end{itemize}

\paragraph{Code search}
The ``Code search'' tool allows searching for file contents in git repository files. It takes simple string and greps the entire codebase. An optional file argument is available to limit search to a specific file. This allows the agent to efficiently locate relevant keywords, functions, files, or code patterns. The output reports all matches along with their corresponding file paths and the line of code in which they occur.

\paragraph{File search}
This tool searches for files in a Git repository by filename, keyword, or glob pattern. It uses 'git ls-files' to list files tracked by the repository and supports glob-style searches. This enables the agent to to obtain a high-level overview of the repository’s structure and contents, or to determine the existence of specific files.

\paragraph{Open file}
A file can be opened by requesting the path to the file. It may also be opened at a specific line. This allows the agent to look at the contents of files. In order to avoid exceeding LLM context limits, only a total of 100 lines is shown at any time.

\paragraph{Scroll file}
Once a file has been opened, the agent can scroll up or down through the file to view the rest of it. It scrolls 100 lines at a time.

\paragraph{Final answer}
Once the agent is ready to answer, it uses the ``Final answer'' tool. This is a default tool of the smolagents \citep{smolagents:1.21.2} library. Once called, it terminates the run-loop and outputs any supplied arguments.

\section{Experimental Design}
\label{chap:experimental-design}

\subsection{Research Questions}
\label{sec:research-questions}

Our study addresses the following research questions:

\begin{itemize}
    \item \textbf{RQ1: How effective is Favia compared with existing vulnerability fix detection approaches?}
    \item \textbf{RQ2: Why do agent-based approaches fail when they make incorrect predictions?}
    \item \textbf{RQ3: How efficient is Favia compared to other approaches?}
\end{itemize}

\subsection{Design to answer RQ1}
\label{sec:rq1-design}
To answer RQ1, we chose to evaluate a diverse set of open-source Large Language Models (LLMs) that differ substantially in scale, architecture, and design goals. 

\subsubsection{Models}
\label{sec:models}

All selected models are instruction-tuned, which is required for their use in agent-based settings, as instruction tuning improves adherence to structured prompts, tool-use conventions, and multi-step task execution. The following models are used:

\begin{itemize}
    \item \textbf{gemma-3-27b-it} \citep{gemmateam2025gemma3technicalreport} is a 27 billion parameter, decoder-only transformer model released by Google as part of the Gemma 3 family. It is instruction-tuned using supervised fine-tuning and alignment techniques to improve instruction following and reasoning behavior. Compared to very large frontier models, gemma-3-27b-it offers a favorable trade-off between model capacity and computational cost, making it suitable for controlled experimental evaluations where inference efficiency is a concern. The model supports long-context inputs of 128K and is trained with a strong emphasis on reasoning, coding, and general-purpose instruction following.

    \item \textbf{Llama-3.3-70B-Instruct} \citep{llama3_3}is a 70 billion parameter dense transformer model developed by Meta. It is instruction-tuned on a mixture of supervised and preference-based data to improve helpfulness and safety. Llama-3.3-70B-Instruct is designed as a high-capacity general-purpose model with strong performance on reasoning, code understanding, and long-context tasks. With its large context window of 128K tokens, its scale and dense architecture serves as a strong open-source baseline for tasks that require deep semantic understanding and robust multi-step reasoning.

    \item \textbf{Qwen3-235B-A22B-Instruct-2507} \citep{qwen3technicalreport} is a Mixture-of-Experts (MoE) instruction-tuned model from the Qwen3 family. The model has a total of 235 billion parameters, with 22 billion parameters active per forward pass. This MoE architecture enables very high representational capacity while keeping inference costs lower than a comparably sized dense model. The instruction-tuned variant is explicitly optimized for complex reasoning, code-related tasks, and agentic workflows involving tool use and decision making. The model has a very large context window of 256K tokens.
\end{itemize}

\subsubsection{Baselines}

We compare Favia against several representative works. As the most commonly used baseline for vulnerability fix detection, we select the machine learning method VulFixMiner \cite{zhou2021finding}. We also select the state-of-the-art (SOTA) machine learning classifier approach PatchFinder \citet{li2024patchfinder:}. Finally, we evaluate against the SOTA LLM-based frameworks LLM4VFD \citet{yang2025code} and CommitShield \cite{wu2025commitshield}.

\textbf{VulFixMiner} by
\cite{zhou2021finding} is a transformer-based approach for automatically identifying silent vulnerability-fixing commits. It analyzes commit-level code changes without relying on commit messages, leveraging a fine-tuned CodeBERT model to learn semantic representations of added and removed code. File-level representations are aggregated into a commit-level embedding, which is then used to rank commits by their likelihood of fixing vulnerabilities.
We base our implementation of VulfixMiner on the implementation by VulCurator \footnote{\url{https://github.com/ntgiang71096/vfdetector/blob/main/vulfixminer.py}.}
. We closely follow VulFixMiners' original model selection and hyperparameters. However, as our training dataset is roughly 100 times larger than that of VulFixMiner, we increase the effective batch size a hundred fold. From 8 to 800 during first training phase, and from 32 to 3200.

\textbf{PatchFinder}
\label{sec:patchfinder}
by \citet{li2024patchfinder:} is a two-phase framework designed to trace security patches for disclosed vulnerabilities in open-source software. PatchFinder first performs an initial retrieval, using both lexical (tf-idf) and semantic (pre-trained code model) similarity to narrow down a candidate set of commits. It then applies a re-ranking phase with an end-to-end learned model to capture deeper semantic correlations between CVE descriptions and commits, enabling more accurate ranking of vulnerability patches.
PatchFinder is open-sourced and available at GitHub\footnote{\url{https://github.com/MarkLee131/PatchFinder}}. Following the original implementation, we use the tf-idf vectorizer from Scikit-Learn \cite{scikit-learn} library to calculate the tf-idf score for the training dataset split. Similarly, we use the CodeReviewer \cite{li2022automatingcodereviewactivities} model from microsoft for calculating the semantic similarity using the CR\_score\footnote{\url{https://github.com/MarkLe e131/CR_score/tree/issta24}}. These two similarities are fused together and keep the top 100 commit candidates for each cve in our dataset. We then fine-tune CodeReviewer to re-rank the top-100 candidate patches from the training dataset. We keep hyperparameters unchanged from the original paper.

\textbf{LLM4VFD} by \citet{yang2025code} is a framework that leverages Large Language Models with structured reasoning and contextual learning to identify vulnerability-fixing commits in open-source software. The framework uses commit intent analysis (CCI), related development artifacts (DA), and historical vulnerability fixes (HV). We use LLM4VFD as a baseline but exclude the development artifacts to ensure a fair comparison, as such artifacts are not available to our agent-based approach. According to the original paper’s ablation analysis, incorporating development artifacts yields a maximum F1-score improvement of only 0.03, indicating that their exclusion has a negligible impact on overall performance.
We base the implementation of LLM4VFD on the replication package provided by the paper \footnote{\url{https://doi.org/10.5281/zenodo.13776994}}. We first construct a vector database of historical vulnerability fixes (HV). We generate 3-aspect summaries for all patching commits in our CVEVC dataset using the three models described in \cref{sec:models}. We create embeddings of these summaries using Qwen/Qwen3-Embedding-8B \citep{zhang2025qwen3embeddingadvancingtext}, and store this in a ChromaDB \citep{chroma:1.0.20} vector database.

\textbf{CommitShield} by \citep{wu2025commitshield} is an LLM-based framework for vulnerability-fixing commit identification. CommitShield operates by enriching commit descriptions and analyzing the patches they contain. It first collects additional information related to the commit description and leverages a large language model (LLM) to generate a more detailed summary. The tool then evaluates the relevance of patches, retaining only those connected to the description. These patches are categorized into intra-procedural (changes within a single function) and inter-procedural (changes involving multiple functions). For intra-procedural patches, CommitShield prepares contextual information about the modified functions, while for inter-procedural patches, it gathers details about related function calls. To support this process, CommitShield employs Joern \citep{joern:1.0.20} to generate Code Property Graphs (CPGs), enabling structured program analysis. Finally, all of this organized information is used to support CommitShield’s vulnerability and fault detection analysis.
While the original work by \citet{wu2025commitshield} only considers C and C++, our dataset is not constrained to a single programming language. We therefore extend CommitShield to support the languages C, C++, Python, PHP, Java, and Go. These languages are facilitated by the main Tree-Sitter\citep{tree-sitter:0.21.3} project, and are also supported by Joern. The languages account for more than 50\% of all files in the diffs of our evaluation datasets. Patches of unsupported languages will be judged based on the description and content of the diff.

\subsubsection{Datasets}

Prior work frequently evaluates vulnerability-fix detection on randomly sampled commits, which may substantially underestimate real-world difficulty. To address RQ1, we evaluate all approaches under two settings: (i) a random setting, where candidate commits are uniformly sampled from the repository, and (ii) a realistic setting, where all commits are potential candidates.

Exhaustively evaluating every commit for each CVE is computationally expensive and not scalable. More importantly, the most challenging cases arise from commits that are highly similar to the true vulnerability fix but do not actually resolve the vulnerability. As discussed in \cref{section:vulnerability-fix-detection}, random sampling used in prior studies is unlikely to include such hard negatives, leading to overly optimistic performance estimates.

To approximate realistic conditions while remaining scalable, Favia first use an efficient machine-learning–based ranking step to filter out commits that are clearly unrelated to the vulnerability. We then evaluate how Favia and the baseline methods identify the true fixing commit among the remaining candidates, which are similar to the correct fix. By comparing performance across these settings, we quantify how random sampling inflates reported performance and assess whether conclusions drawn from simplified evaluations hold under realistic deployment conditions.

Our raw data comes from previous work \cite{sun2024where}, which contains 17,293 CVEs with 4,682 corresponding open-source repositories and 23,303 patching commits. 
The patching commits span over 200 different file types, including different languages like Java (13.7\%), PHP (13.6\%), C (13.1\%), JavaScript (10.0\%), Python (4.3\%), Go (4.1\%), and others (41.2\%), which reflects the complicated and highly unbalanced code environments for vulnerability fix tracing.

For every CVE in the dataset, we git clone every repository that is a GitHub repository. We get all patch commits, and select up to 5,000 non-patch commits at random. If a repository has fewer than or equal to 5,000 commits, we select all. In total, we identify a total of 3,820 repositories, of which 3,708 are successfully downloaded, resulting in approximately 2 terabytes of data. We also try to remove binary or otherwise excessively large diffs from the non-patch set; we filter commits based on diff size. These are primarily dependency updates, vendor code imports, or asset changes that are unlikely to represent vulnerability fixes and would otherwise add noise and unnecessary computational overhead. Specifically, we retain diffs up to the 95th percentile of diff string length (measured in number of characters); any diff exceeding 153,993 characters is discarded. After filtering, a total of 8,283,424 commits are left. We split the patch commits into training, validation, and test sets using an 80/10/10 split. To prevent data contamination, the split is performed at the repository level, ensuring that no repository appears in more than one split. Non-patch commits are then assigned to splits based on the repository-level partitioning induced by the patch data. We name the collected dataset CVEVC, short for Common Vulnerability Enumeration Vulnerability-fixing Commits. Due to the large amount of data, we normalize the dataset into three separate datasets. One dataset stores unique commit data \footnote{\url{https://huggingface.co/datasets/andstor/cvevc_commits}}, another stores CVE data \footnote{\url{https://huggingface.co/datasets/andstor/cvevc_cve}}, and the third dataset stores mapping between labeled commits to CVEs \footnote{\url{https://huggingface.co/datasets/andstor/cvevc_cve_commit_mappings}}.

From the CVEVC dataset, we construct two new datasets \footnote{\url{https://huggingface.co/datasets/andstor/cvevc_candidates}}. One based on random sampling, which is based on the dataset construction strategies of baselines \citep{zhou2021finding, li2024patchfinder:, yang2025code}. Another dataset emulates a realistic scenario in which all commits are considered candidates. However, to perform scalable comparisons between Favia and baselines, we focused on the most challenging aspect of identifying the correct commit in the realistic dataset. We assume that the top candidates (the top 10 in our design) are most likely the correct ones identified. We then compare how well the tools can find the correct one from the top candidates. 

\textbf{Random dataset}. \label{sec:random-dataset} We construct the ``random\_10'' evaluation dataset consisting of randomly selected commits from the test split. For each CVE, we select up to 10 commits. This includes all corresponding patch commits and randomly samples the rest with non-patch commits from the same repository. If fewer than 10 non-patch commits are available, all are included. The sampled non-patch commits may include other patching commits that are unrelated to the specific CVE under consideration.

\begin{figure}[htb]
    \centering
    \includegraphics{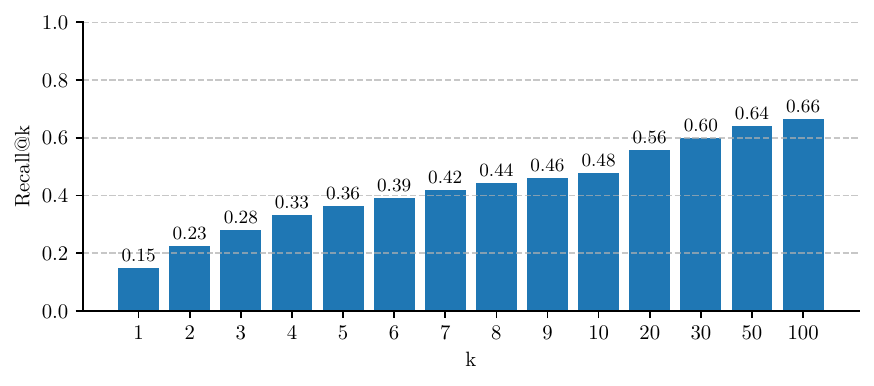}
    \caption{PatchFinder effectiveness on test split of CVEVC dataset.}
    \label{fig:patchfinder-recall}
\end{figure}

\textbf{Realistic dataset}. \label{sec:realistic-dataset} To identify the top 10 ``most difficult'' commit candidates, we use PatchFinder, trained on the training split of the CVEVC dataset (see \cref{sec:patchfinder}) to rank the commits for each CVE in the test split of the dataset. \Cref{fig:patchfinder-recall} shows the distribution of recall@k, computed over the top-k ranked commits across all available patches. We select the top 10 ranked commits for each CVE, and package this as ``PatchFinder\_top10''. The selected commits may very well include other patching commits that are unrelated to the specific CVE under consideration. As we select \(k=10\), any downstream evaluation on this dataset can at most find 48\% of the total patches. 

\subsubsection{Metrics}
We evaluate vulnerability fix detection as a binary classification task at the commit level, where each candidate commit is classified as either a vulnerability-fixing patch or a non-patch commit for a given CVE. Due to the highly imbalanced nature of the task, we report precision, recall, and F1-score. This comparison allows us to determine whether agent-based, iterative reasoning provides measurable benefits over current state-of-the-art (SOTA) methods. All methods are evaluated on the same candidate sets using identical train/validation/test splits and metrics.

\textbf{Precision} measures the proportion of commits predicted as vulnerability fixes that are correct. High precision indicates that a method produces few false positives, which is critical in large repositories where incorrectly labeling benign commits as security patches can lead to unnecessary manual inspection or incorrect downstream actions.

\textbf{Recall} measures the proportion of true vulnerability-fixing commits that are successfully identified. High recall is essential in security settings, as missing a true patch may result in an unmitigated vulnerability remaining in the codebase or downstream dependencies.

\textbf{F1-score} is the harmonic mean of precision and recall, providing a single metric that balances false positives and false negatives. It is particularly useful when comparing methods with different trade-offs between precision and recall on imbalanced datasets.

All metrics are computed by aggregating predictions across all CVEs within a dataset split, reflecting a realistic deployment scenario where a system must identify vulnerability-fixing commits among a large number of unrelated changes. Accuracy is omitted, as it would be dominated by the majority non-patch class and thus provide limited insight.

\subsection{Design to answer RQ2}
\label{sec:rq2-design}
To understand the limitations of agent-based reasoning, RQ2 analyzes failure cases produced by Favia. We examine agent trajectories, tool usage patterns, and reasoning traces, and categorize incorrect predictions into distinct failure modes. This analysis reveals whether failures stem from insufficient exploration, misunderstanding of vulnerability semantics, overconfidence, or reliance on surface-level cues, providing insight into how agent behavior can be improved.

\subsubsection{Agent trajectories}
From traces collected from experiments of RQ1, we analyze the amount of calls to the different tools. Because a model can invoke arbitrary many tool calls at any one step, this might give misleading information. For example, opening a file and scrolling in a loop with 100 iterations. We therefore only count invocation of each tool type once per step.

\subsubsection{Failure mode classification}

We categorize incorrect classifications by prompting an Large Language Model (LLM) to classify the agent's run history into eight failure modes defined in \cref{tab:failure-mode-categories}.

\begin{table}[htbp]
    \centering
    \newcolumntype{Y}{>{\centering\arraybackslash}X}
    \newcolumntype{R}{>{\raggedright\arraybackslash}X}
    \newcolumntype{L}{>{\raggedleft\arraybackslash}X}
    \centering
    \caption{Failure mode categories.}\label{tab:failure-mode-categories}
    \begin{tabularx}{\textwidth}{l|X}
        \toprule
        \textbf{Category} & \textbf{Description} \\
        \midrule
        Superficial Association & The model inferred a match due to keywords, filenames, or conceptual similarity without real evidence. \\
        \midrule
        Failed to Find Relevant Context & The agent were not able to find the relevant files. \\
        \midrule
        CVE Misinterpretation & The model misinterpreted the CVE’s root cause, vulnerable component, or exploit mechanism—resulting in a justification that does not align with the actual CVE description. \\
        \midrule
        Memorized CVE & The model did not retreve the CVE report and concluded based on memory.\\
        \midrule
        Incorrect Classification & The agent collected reasonable evidence but came to the wrong conclusion.\\
        \midrule
        Ran Out Of Budget & The agent seem to be on the right track, but the episode ended prematurely. \\
        \midrule
        Gave Up Prematurely & The agent decided to stop solving the problem after encountering some difficulty. \\
        \midrule
        Other & There was some other problem that prevented the agent from correctly classifying the commit. \\
        \bottomrule
    \end{tabularx}
\end{table}

In order to provide a fair and unbiased evaluation, we use a LLM from an independent model family, separate from the models in \cref{sec:models}. Specifically, we use gpt-oss-120b \citep{openai2025gptoss120bgptoss20bmodel}. gpt-oss-120b is an open-weight Mixture-of-Experts model released by OpenAI, featuring approximately 117 billion total parameters with about 5.1 billion active parameters per token, designed for general-purpose and high-reasoning tasks. It supports configurable reasoning effort and supports a 128K tokens context window.

We collect all Favia's agent traces from the results of RQ1. We then filter out all traces that were correctly classified. We then prompt the LLM to classify Favia's agent traces. In the classification prompt, we supply the CVE description and the agent trace. In order to focus the analysis on the agent behavior, in the agent trace, we only include the agent task, reasoning steps, and tool calls. We exclude any tool outputs. We configure gpt-oss-120b to use a medium reasoning effort.

\subsection{Design to answer RQ3}
\label{sec:rq3-design}
Agent-based reasoning introduces additional computational cost due to multi-step interaction and context accumulation. To answer RQ3,  we measure efficiency in terms of input, output, and embedding token consumption across all approaches. We report mean token usage per commit and analyze how cost scales with model size and reasoning depth. This allows us to characterize the trade-off between improved detection performance and increased computational overhead, and to assess the practicality of deploying agent-based systems in large-scale settings.

\section{Experimental Results}
\label{chap:experimental-results}

In this section, we present detailed results of each research question. 

\subsection{Results of RQ1: Effectiveness}
\label{sec:effectiveness}

\subsubsection{Results based on the random dataset}

Table \ref{tab:eval-summary-random10} reports performance of Favia and the baselines on the \textbf{random dataset}, where candidate commits are sampled uniformly at random, using the dataset construction strategies in +++ref. Considering performance ranges across models provides a more robust comparison than focusing on individual model–approach pairs.

The results show that \textbf{VulFixMiner} and \textbf{PatchFinder} benefit from random sampling primarily in terms of precision. VulFixMiner reaches a precision of 0.83 but continues to exhibit extremely low recall (0.01), resulting in a negligible F1-score. PatchFinder achieves high precision (0.86) and moderate recall (0.23), improving its F1-score to 0.36. Despite these gains, both methods remain recall-limited, even in this simplified setting.
LLM-based techniques, i.e., \textbf{LLM4VFD} and \textbf{CommittShield}, show the largest absolute gains under random evaluation. Across models, precision ranges from 0.47 to 0.82, while recall remains consistently high (0.71–0.94), yielding F1-scores of up to 0.87. 
\textbf{Favia} achieves the strongest overall performance across models, with precision between 0.59 and 0.82, recall between 0.86 and 0.93, and F1-scores ranging from 0.72 to 0.87.

\begin{table*}[htbp]
\begin{threeparttable}
    \newcolumntype{Y}{>{\centering\arraybackslash}X}
    \newcolumntype{R}{>{\raggedright\arraybackslash}X}
    \newcolumntype{L}{>{\raggedleft\arraybackslash}X}
    \centering
    \caption{Performance metrics across different approaches on Random 10 subset}\label{tab:eval-summary-random10}
    \begin{tabularx}{\textwidth}{ll!{\color{white}\hspace{.5em}}YYY}
        \toprule
        \textbf{Model} & \textbf{Approach} & \textbf{Precision} & \textbf{Recall} & \textbf{F1-score} \\
        \midrule
        \multirow{1}{*}{CodeBERT} & VulFixMiner & 0.83 & 0.01 & 0.03 \\
        \midrule
        \multirow{1}{*}{CodeReviewer} & PatchFinder & 0.86 & 0.23 & 0.36 \\
        \midrule
        \multirow{3}{*}{Llama-3.3-70B-Instruct} & LLM4VFD & \textbf{0.77} & 0.74 & 0.75 \\
        & CommitShield & 0.45 & 0.75 & 0.56 \\
        & Favia & 0.74 & \textbf{0.86} & \textbf{0.79} \\
        \midrule
        \multirow{3}{*}{Qwen3-235B-A22B-Instruct-2507} & LLM4VFD & 0.74 & 0.81 & 0.77 \\
        & CommitShield & 0.62 & 0.71 & 0.66 \\
        & Favia & \textbf{0.82} & \textbf{0.92} & \textbf{0.87} \\
        \midrule
        \multirow{3}{*}{gemma-3-27b-it} & LLM4VFD & 0.47 & 0.87 & 0.61 \\
        & CommitShield & 0.29 & \textbf{0.94} & 0.44 \\
        & Favia & \textbf{0.59} & 0.93 & \textbf{0.72} \\
       \bottomrule
    \end{tabularx}
    \begin{tablenotes}[flushleft]\small
      \item \textbf{Bold}: best-performing approach per model.
    \end{tablenotes}
\end{threeparttable}
\end{table*}

\subsubsection{Results based on the realistic dataset}

Table \ref{tab:eval-summary-patchfindertop10} highlights clear performance strata between traditional methods, LLM-based frameworks, and the proposed agent-based approach (Favia) on the realistic dataset. 

\begin{table*}[htbp]
\begin{threeparttable}
    \newcolumntype{Y}{>{\centering\arraybackslash}X}
    \newcolumntype{R}{>{\raggedright\arraybackslash}X}
    \newcolumntype{L}{>{\raggedleft\arraybackslash}X}
    \centering
    \caption{Performance metrics across different approaches on the realistic dataset}\label{tab:eval-summary-patchfindertop10}
    \begin{tabularx}{\textwidth}{ll!{\color{white}\hspace{.5em}}YYY}
        \toprule
        \textbf{Model} & \textbf{Approach} & \textbf{Precision} & \textbf{Recall} & \textbf{F1-score} \\
        \midrule
        \multirow{1}{*}{CodeBERT} & VulFixMiner & 0.43 & 0.03 & 0.06 \\
        \midrule
        \multirow{1}{*}{CodeReviewer} & PatchFinder & 0.40 & 0.37 & 0.38 \\
        \midrule
        \multirow{3}{*}{Llama-3.3-70B-Instruct} & LLM4VFD & \textbf{0.30} & 0.89 & 0.45 \\
        & CommitShield & 0.18 & 0.87 & 0.30 \\
        & Favia & \textbf{0.30} & \textbf{0.94} & \textbf{0.46} \\
        \midrule
        \multirow{3}{*}{Qwen3-235B-A22B-Instruct-2507} & LLM4VFD & 0.29 & 0.93 & 0.45 \\
        & CommitShield & 0.22 & 0.85 & 0.35 \\
        & Favia & \textbf{0.39} & \textbf{0.98} & \textbf{0.56} \\
        \midrule
        \multirow{3}{*}{gemma-3-27b-it} & LLM4VFD & 0.18 & 0.94 & 0.30 \\
        & CommitShield & 0.12 & \textbf{0.99} & 0.21 \\
        & Favia & \textbf{0.23} & 0.98 & \textbf{0.37} \\
       \bottomrule
    \end{tabularx}
    \begin{tablenotes}[flushleft]\small
      \item \textbf{Bold}: best-performing training approach per model.
    \end{tablenotes}
\end{threeparttable}
\end{table*}

Regarding each evaluated approach, \textbf{VulFixMiner} exhibits high precision but extremely low recall, with precision around 0.43 and recall as low as 0.03, yielding F1-scores in the 0.06 range. This narrow operating regime reflects a highly conservative detector: when it predicts a vulnerability-fixing commit it is often correct, but it fails to recover almost all true patches. As a result, VulFixMiner performs poorly in recall-oriented security settings and does not scale well to large, diverse candidate sets. \textbf{PatchFinder} shows a more balanced but still limited performance profile, with precision and recall both lying in the 0.37–0.40 range and an F1-score of approximately 0.38. Compared to VulFixMiner, PatchFinder substantially improves recall while maintaining competitive precision. However, its performance remains bounded by similarity-based retrieval and re-ranking, which limits its ability to capture deeper semantic and contextual relationships present in complex patches.  

For the LLM-based techniques, \textbf{LLM4VFD} consistently achieves very high recall, ranging from 0.89 to 0.94, but with low precision, ranging from 0.18 to 0.30. This results in F1-scores between 0.30 and 0.45. These results indicate that LLM4VFD is effective at identifying most vulnerability-fixing commits but tends to over-predict, leading to many false positives. Increasing model capacity improves recall slightly but does not substantially close the precision gap. \textbf{CommitShield} follows a similar but slightly weaker trend compared to LLM4VFD. Recall ranges from 0.85 to 0.99, while precision drops to 0.12–0.22, producing F1-scores between 0.21 and 0.35. The method is strongly recall-oriented, benefiting from structural program analysis, but its aggressive inclusion of candidate patches results in the lowest precision among LLM-based baselines.

\textbf{Favia} consistently delivers the strongest and most stable performance across models. Precision improves substantially compared to other LLM-based methods, ranging from 0.23 to 0.39, while recall remains extremely high at 0.94–0.98. This leads to F1-scores between 0.37 and 0.56, representing the best overall range among all approaches. Notably, Favia maintains high recall comparable to LLM4VFD and CommitShield, while significantly reducing false positives, especially with larger models.

\subsubsection{Performance differences using random vesus realistic datasets}

Compared to performance using the random dataset for evaluation (as shown in \ref{tab:eval-summary-random10}), \textbf{\textit{all approaches achieve substantially lower precision, recall, and F1-scores}}, indicating that random evaluation leads to a markedly easier detection task, meaning random sampling significantly reduces ambiguity between vulnerability-fixing and non-fixing commits.  

However, an evident pattern emerges: \textbf{\textit{the performance advantage of Favia over other LLM‑based baselines is more pronounced when using the realistic dataset than when using the randomly generated dataset}}. This highlights the benefits of Favia’s structured, multi‑step reasoning approach. In contrast, this advantage becomes less noticeable when the candidate set is constructed randomly.

Randomly selected commit sets contain a high proportion of trivially non-security-related changes, making vulnerability-fixing commits easier to identify. This substantially inflates precision and F1-scores for all approaches and reduces the relative difficulty of the task. As a result, random evaluation does not accurately reflect real-world deployment scenarios, where all commits are candidates, and most security-relevant commits will have high chance of being classified as false positive.

An additional factor influencing the observed recall differences is the construction of the realistic candidate set. In the realistic setting, evaluation is conditioned on PatchFinder ranking the true vulnerability-fixing commit within the top-10 candidates; if the true patch is not retrieved at this stage, it is excluded from downstream evaluation. Consequently, only 1,081 out of 1,857 true patches are present in the realistic dataset, with the remaining 776 patches omitted. Importantly, the retained patches are not easier cases: they are typically harder, more ambiguous fixes that survive similarity-based ranking and are surrounded by highly similar, security-relevant non-patch commits. However, because each CVE contributes at most one true patch and competing positives unrelated to the target CVE are largely absent, recall is measured over a smaller and more constrained set of positives. In contrast, the random dataset may include multiple patch commits per CVE that address different vulnerabilities and are correctly rejected by Favia due to lack of causal alignment, but are nevertheless counted as false negatives. This difference in evaluation conditioning explains why recall is higher on the realistic dataset despite its greater semantic difficulty, and further underscores the importance of realistic candidate selection when assessing vulnerability-fix detection systems.

\subsubsection{Analysis of Performance differences between Favia and the baselines}
\label{sec:case-study}
To explain Favia's superiour precision, we present a comparison between the results of Favia's, CommitShield's, and LLM4VFD's assessments of the commit 705a427 from the VLC media player (videolan/vlc), as the patch for \href{https://nvd.nist.gov/vuln/detail/cve-2014-9625}{CVE-2014-9625}. The commit 705a427 is ranked number 6 in the top 10 candidate commits selected by PatchFinder for CVE-2014-9625. This is not the correct patch, but is somewhat similar to the correct patch commit. The CVE describes an integer truncation vulnerability in the \texttt{GetUpdateFile} function in \texttt{misc/update.c}, where an incorrect cast from a 64-bit integer to a 32-bit integer during memory allocation for a downloaded update file can result in a buffer overflow, allowing remote attackers to execute arbitrary code via a crafted update status file.

\begin{center}
\bigskip
\begin{tcblisting}{
    title=CommitShield,
    enhanced,
    colbacktitle=red!5!white,
    coltitle=black,
    colback=white,
    colframe=red!75!black,
    boxrule=0.5mm,
    fonttitle=\bfseries,
    fontupper=\ttfamily,
    arc=1mm,
    breakable=true,
    listing only,
    listing options={
        numbers=none,
        frame=none,
        xleftmargin=0pt,
        xrightmargin=0pt,
        columns=flexible,
        breaklines=true,
        breakautoindent=false,
        breakindent=0em,
        lineskip=0pt,
        breakatwhitespace=true,
        escapechar=§,
        literate={\{}{\{}{1}%
                {\}}{\}}{1}%
      }
}
The commit description indicates that the patch checks the return value of `vlc_memstream_open` in the function `vlc_xml_encode` to ensure the stream handle is valid before use, preventing §\lstbg{yellow!40}{potential undefined behavior}§ due to failed stream initialization. This implies that prior to the patch, there was a known issue where the code did not validate the success of stream initialization, which could lead to undefined behavior - a §\lstbg{yellow!40}{runtime error with potential security}§ §\lstbg{yellow!40}{implications}§. The modification adds a check on the return value and returns `NULL` on failure, thus fixing a known vulnerability related to improper error handling. Since the patch addresses a known §\lstbg{yellow!40}{runtime error that could result in}§ §\lstbg{yellow!40}{undefined behavior}§ (a recognized class of security vulnerabilities)}, and the §\lstbg{yellow!40}{language in the description ('prevent potential undefined behavior')}§ suggests a concrete defect was corrected, this §\lstbg{yellow!40}{qualifies as a vulnerability repair commit}§ under the given definition.
\end{tcblisting}
\captionof{figure}{Reasoning output of CommitShield's analysis of commit 705a427 against CVE-2014-9625.}
\label{fig:commit-shield-example}
\bigskip
\end{center}

As shown in \cref{fig:commit-shield-example}, CommitShield incorrectly classifies the commit as a vulnerability patch by over-interpreting generic safety language in the commit description. In particular, its conclusion hinges on the highlighted phrases ``potential undefined behavior'' and ``runtime error with potential security implications'', which CommitShield treats as sufficient evidence of a security fix. This reasoning culminates in the explicit assertion that the change ``qualifies as a vulnerability repair commit'', despite the absence of any linkage to the CVE itself.

However, the highlighted emphasis on undefined behavior reflects only a general robustness improvement, not remediation of the specific vulnerability described by CVE-2014-9625. The CVE concerns an integer truncation leading to a buffer overflow in the \texttt{GetUpdateFile} function in \texttt{misc/update.c}, whereas the commit modifies \texttt{vlc\_xml\_encode} in \texttt{src/text/strings.c}. By relying on surface-level indicators, such as the presence of defensive checks and language suggesting safety improvements, CommitShield conflates generic error handling with CVE-specific vulnerability remediation, leading to a false positive.

\begin{center}
\bigskip
\begin{tcblisting}{
    title=LLM4VFD,
    enhanced,
    colbacktitle=red!5!white,
    coltitle=black,
    colback=white,
    colframe=red!75!black,
    boxrule=0.5mm,
    fonttitle=\bfseries,
    fontupper=\ttfamily,
    arc=1mm,
    breakable=true,
    listing only,
    listing options={
        numbers=none,
        frame=none,
        xleftmargin=0pt,
        xrightmargin=0pt,
        columns=flexible,
        breaklines=true,
        breakautoindent=false,
        breakindent=0em,
        lineskip=0pt,
        breakatwhitespace=true,
        escapechar=§,
        literate={\{}{\{}{1}%
                {\}}{\}}{1}%
      }
}
The current patch modifies the `vlc_xml_encode` function to check the return value of `vlc_memstream_open`, which is a function that can fail due to memory allocation or system resource limitations. Previously, the return value was ignored, and the code proceeded under the assumption that the stream was successfully initialized. This could lead to the use of an invalid or uninitialized `vlc_memstream_t` handle in subsequent operations, potentially resulting in §\lstbg{yellow!40}{undefined behavior such as memory corruption, crashes, or data}§ §\lstbg{yellow!40}{integrity issues}§. By adding a check and returning `NULL` on failure, the patch ensures that the function fails gracefully instead of proceeding with an invalid state.

Comparing this to the §\lstbg{yellow!40}{historical vulnerability fix in Kodi}§, both changes focus on improving robustness by properly handling resource initialization failures. In the Kodi case, improper handling of input streams during XML parsing led to a buffer overflow and denial of service. Similarly, the current patch addresses improper handling of stream initialization, which, while not directly involving buffer overflow, mitigates a path where failure to check a critical function return value could lead to undefined behavior-§\lstbg{yellow!40}{often a precursor to exploitable}§ §\lstbg{yellow!40}{conditions in memory-sensitive contexts}§.

Although the current patch does not address a buffer overflow per se, it corrects a missing error check in a §\lstbg{yellow!40}{security-critical function that handles string}§ §\lstbg{yellow!40}{encoding for XML}§, which is often part of input processing pipelines. Such functions are common targets for attackers when error paths are not properly handled. The fact that the patch prevents use of an uninitialized resource and aligns with secure coding practices (e.g., CERT ERR33-C) suggests it is addressing a potential vulnerability, specifically a resource initialization failure that §\lstbg{yellow!40}{could be leveraged in denial-of-service attacks or lead to}§ §\lstbg{yellow!40}{memory-related vulnerabilities under certain conditions}§.

The commit message emphasizes that the function is 'not guaranteed to succeed,' and using the handle without checking is explicitly acknowledged as unsafe. This indicates awareness of a defect that could compromise program reliability and safety. While no direct exploit is demonstrated, the nature of the fix-adding essential error checking to prevent use of an invalid resource-aligns with vulnerability fixes seen in similar codebases.

Therefore, §\lstbg{yellow!40}{the patch is best classified as a vulnerability fix}§, as it mitigates a potential security issue rooted in improper error handling that could lead to undefined behavior in a function likely used during untrusted input processing.
\end{tcblisting}

\captionof{figure}{Reasoning output of LLM4VFD's analysis of commit 705a427 against CVE-2014-9625.}
\label{fig:llm4vfd-example}
\bigskip
\end{center}

LLM4VFD similarly misclassifies the commit as a vulnerability fix, as shown in \cref{fig:llm4vfd-example}, but does so through speculative analogy rather than explicit CVE alignment. Its reasoning repeatedly emphasizes highlighted phrases such as ``undefined behavior'', ``memory corruption'', and ``often a precursor to exploitable conditions'', framing the change as security-relevant despite acknowledging that it ``does not address a buffer overflow per se.''

The model further reinforces this conclusion by drawing comparisons to a ``historical vulnerability fix in Kodi'' and by labeling \texttt{vlc\_xml\_encode} as a ``security-critical function'', even though no such claim is made in the CVE description. These highlighted analogies substitute pattern-based reasoning for concrete evidence. Crucially, LLM4VFD never references the integer truncation, the 64-bit to 32-bit cast, or the GetUpdateFile function central to CVE-2014-9625. As a result, the final highlighted conclusion “the patch is best classified as a vulnerability fix” rests on abstract security heuristics rather than CVE-grounded causal alignment, again producing a false positive.

\begin{center}
\bigskip
\begin{tcblisting}{
    title=Favia,
    enhanced,
    colbacktitle=green!5!white,
    coltitle=black,
    colback=white,
    colframe=green!50!black,
    boxrule=0.5mm,
    fonttitle=\bfseries,
    fontupper=\ttfamily,
    arc=1mm,
    breakable=true,
    listing only,
    listing options={
        numbers=none,
        frame=none,
        xleftmargin=0pt,
        xrightmargin=0pt,
        columns=flexible,
        breaklines=true,
        breakautoindent=false,
        breakindent=0em,
        lineskip=0pt,
        breakatwhitespace=true,
        escapechar=§,
        literate={\{}{\{}{1}%
                {\}}{\}}{1}%
      }
}
The CVE-2014-9625 vulnerability is an §\lstbg{yellow!40}{integer truncation issue}§ in the §\lstbg{yellow!40}{GetUpdateFile function in src/misc/update.c}§, related to a 64-bit to 32-bit cast during memory allocation for a downloaded update file. The commit 705a427f75915d9951f62fa86851693dc16a73a7, however, modifies §\lstbg{yellow!40}{vlc\_xml\_encode in src/text/strings.c}§ by adding a check for the return value of vlc_memstream_open. This change is unrelated to §\lstbg{yellow!40}{integer casting, buffer overflow from truncation, or the update mechanism}§. The affected component, function, and vulnerability type do not match. Therefore, this commit is not a patch for CVE-2014-9625.
\end{tcblisting}

\captionof{figure}{Reasoning output of Favia's analysis of commit 705a427 against CVE-2014-9625.}
\label{fig:favia-example}
\bigskip
\end{center}

Favia correctly identifies the commit as a non-patch, as shown in \cref{fig:favia-example}, by grounding its analysis in explicit alignment between the CVE description and the code changes. Favia first extracts the vulnerability semantics from the CVE—namely, an integer truncation during memory allocation in \texttt{GetUpdateFile} within \texttt{misc/update.c}—and then systematically verifies whether the commit addresses the same component, function, and vulnerability type. Since commit 705a427 exclusively modifies XML string encoding logic in \texttt{src/text/strings.c} and introduces no changes related to integer casting, buffer sizing, or the update mechanism, Favia correctly concludes that the commit is unrelated. Upon manually analyzing the full trace of Favia\footnote{\url{https://andstor-phoenix-cvevc-candidates-patchfinder-top10.hf.space/projects/UHJvamVjdDoz/traces/c21e04f502c97d292247a7fc63c8ba05}}, we also see the agent tries to find the \texttt{GetUpdateFile} inside \texttt{misc/update.c} and finds it is already fixed. This flexible evidence-driven, component-aware reasoning enables Favia to avoid false positives that arise from generic security heuristics, demonstrating its superior precision compared to CommitShield and LLM4VFD.

Together, these outcomes illustrate why Favia provides the most accurate assessment: by requiring semantic, structural, and causal alignment between CVE descriptions and code changes, it distinguishes true vulnerability patches from incidental robustness improvements, significantly reducing false positive classifications.

\vspace{1em}
\noindent\fbox{%
    \parbox{\columnwidth}{%
        \textbf{Summary for RQ1:} Favia is more effective than state-of-the-art techniques for vulnerability-fixing commit identification. Compared to traditional methods, it avoids the severe recall limitations, and compared to existing LLM-based approaches, it substantially reduces false positives while preserving very high recall. LLM-based vulnerability fix detection appears significantly more effective when evaluated on randomly sampled commits than under realistic candidate selection. Random evaluation inflates precision and F1-scores across all approaches and narrows the performance gap between competing methods. In contrast, PatchFinder-based evaluation better reflects real-world conditions and more clearly reveals the strengths of agent-based reasoning.
    }%
}

\subsection{Results of RQ2: Failure analysis results}

As explained in \cref{sec:rq2-design}, we analyzed agent trajectories, tool usage patterns, and reasoning traces, and categorized incorrect predictions into distinct failure modes. In the following sections, we characterize some of the common system behaviors, including the common patterns arising from the order and volume of tool call usage, and the summary of failure modes and their possible reasons.

\subsubsection{Results of trajectory analysis}
\begin{figure}[htbp]
    \centering
    \includegraphics[width=\textwidth]{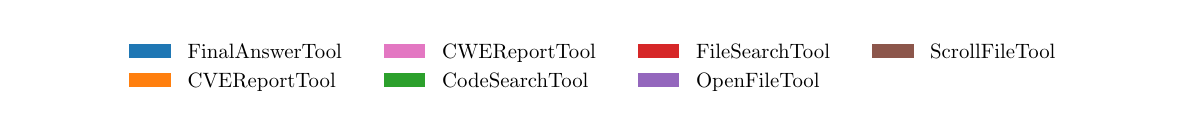}
    \hfill
    \begin{subfigure}[t]{0.37333\textwidth}
        \centering
        \includegraphics[width=\textwidth]{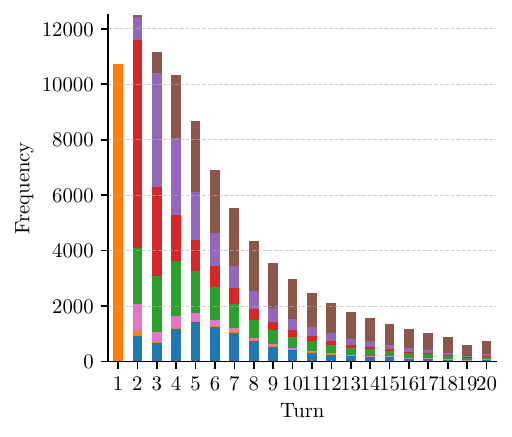}
        \caption{Qwen3-235B-A22B-Instruct-2507}
        \label{fig:tool-calls-realistic-qwen}
    \end{subfigure}%
    \hfill
    \begin{subfigure}[t]{0.31\textwidth}
        \centering
        \includegraphics[width=\textwidth]{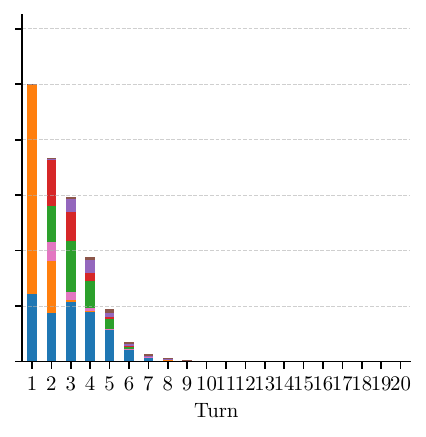}
        \caption{Llama-3.3-70B-Instruct}
        \label{fig:tool-calls-realistic-llama}
    \end{subfigure}%
    \hfill
    \begin{subfigure}[t]{0.31\textwidth}
        \centering
        \includegraphics[width=\textwidth]{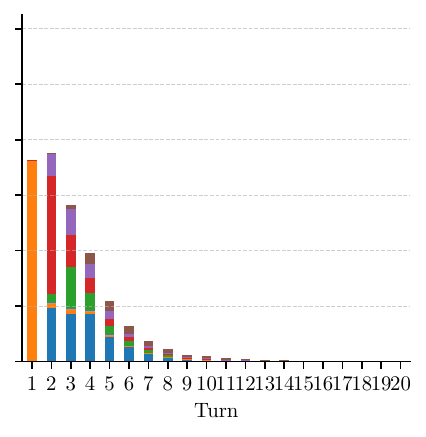}
        \caption{gemma-3-27b-it}
        \label{fig:tool-calls-realistic-gemma}
    \end{subfigure}%
    \caption{The frequency of tool calls invoked at each turn on the realistic dataset.}
    \label{fig:tool-calls-realistic}
\end{figure}

\begin{figure}[htbp]
    \centering
    \includegraphics[width=\textwidth]{figures/tool_calls_legend.pdf}
    \hfill
    \begin{subfigure}[t]{0.37333\textwidth}
        \centering
        \includegraphics[width=\textwidth]{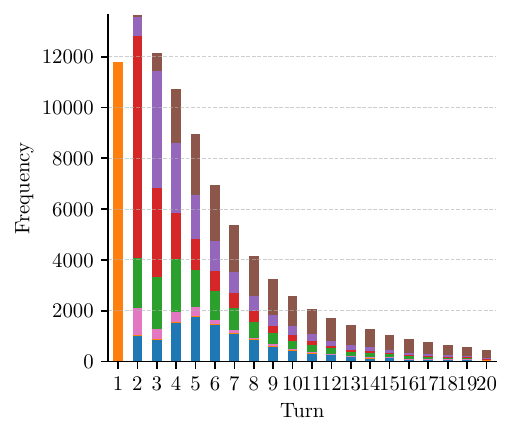}
        \caption{Qwen3-235B-A22B-Instruct-2507}
        \label{fig:tool-calls-random-qwen}
    \end{subfigure}%
    \hfill
    \begin{subfigure}[t]{0.31\textwidth}
        \centering
        \includegraphics[width=\textwidth]{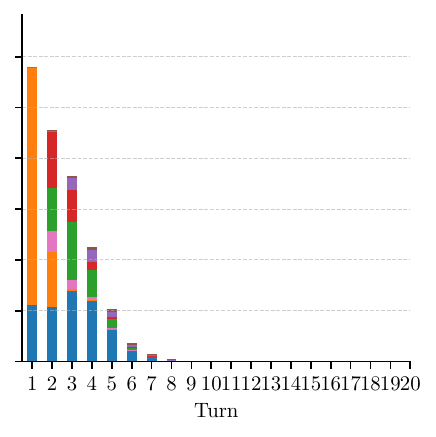}
        \caption{Llama-3.3-70B-Instruct}
        \label{fig:tool-calls-random-llama}
    \end{subfigure}%
    \hfill
    \begin{subfigure}[t]{0.31\textwidth}
        \centering
        \includegraphics[width=\textwidth]{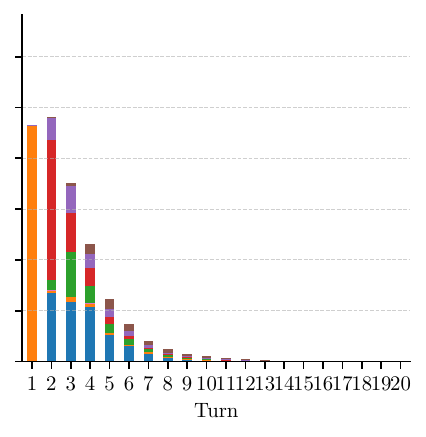}
        \caption{gemma-3-27b-it}
        \label{fig:tool-calls-random-gemma}
    \end{subfigure}%
    \caption{The frequency of tool calls invoked at each turn on the random dataset.}
    \label{fig:tool-calls-random}
\end{figure}

The results of the trajectory analyses are shown in \cref{fig:tool-calls-realistic} and \cref{fig:tool-calls-random}. The observed trends and patterns are characterized in the following paragraphs.

\paragraph{Low degree of memorization.} Nearly all successful trajectories begin with an explicit call to the CVEReportTool. The CVE description is imperative, as it provides the task definition. As described in \cref{sec:tools-cve-report}, omitting this call implies that the model proceeds without externally retrieving the CVE description, suggesting reliance on internalized knowledge from pretraining and thus a potential short-circuit of the intended reasoning process.

We observe substantial variation across models. Qwen consistently invokes the CVEReportTool in the first turn across both datasets, with no successful trajectories starting directly with the FinalAnswerTool. In contrast, on the realistic dataset, Llama and Gemma exhibit 2,958 and 61 rounds, respectively, where the FinalAnswerTool is called without any prior CVE retrieval. Similarly, on the random dataset, Llama and Gemma exhibit 2,421 and 46 rounds, respectively. This suggests that these models rely more heavily on memorized patterns or heuristics, bypassing explicit evidence retrieval and thereby reducing the faithfulness of the intended reasoning process.

Despite this, the overall frequency of such memorized cases remains relatively low relative to the total number of successful attempts. For Qwen, Llama, and Gemma, the corresponding rates are 0\%, 24\%, and 0.8\% on the realistic dataset, and 0\%, 21\%, and 0.5\% on the random dataset. Upon manually inspecting the potentially memorized runs, many of the cases are credited due to changes being only documentation, or completely irrelevant code changes. Hence, the agent dismisses these as not possibly related to any vulnerability and calls FinalAnswerTool without needing to check the CVE report.

\paragraph{Early localization through file search.}
After retrieving the CVE description, the second turn is dominated by localization actions. As shown in \cref{fig:tool-calls-realistic} and \cref{fig:tool-calls-random}, the most frequently invoked tool is the \texttt{FileSearchTool}, primarily used to identify files explicitly mentioned in the CVE report or implied by the affected components. This step grounds the abstract vulnerability description in concrete locations within the repository, enabling targeted inspection of relevant code regions.

\paragraph{Progressive narrowing through code navigation.}
Subsequent turns typically follow a structured exploration pattern, where the agent alternates between opening files, scrolling through relevant code regions, and inspecting related functions or call sites. This repeated open--read--navigate loop reflects a progressive narrowing of focus from repository-level context to specific code changes, allowing the agent to align the commit diff with the vulnerable logic described in the CVE.

\subsubsection{Results of failure analyses}
We categorize incorrect classifications by prompting gpt-oss-120b to classify the agent's run history into eight failure modes defined in \cref{tab:failure-mode-categories}. See \cref{sec:rq2-design} for detailed steps.

\begin{figure*}[t]
    \centering
    \begin{subfigure}[t]{0.5\textwidth}
        \centering
        \includegraphics[scale=0.9]{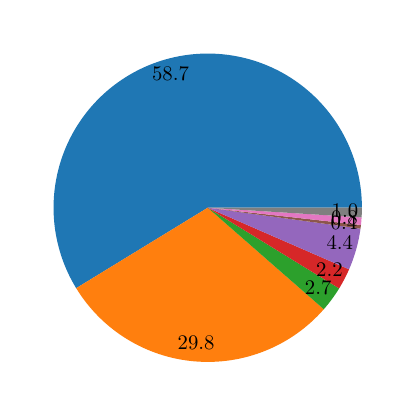}
        \caption{Realistic dataset.}
        \label{fig:failure-modes-patchfinder}
    \end{subfigure}%
    \hfill
    \begin{subfigure}[t]{0.5\textwidth}
        \centering
        \includegraphics[scale=0.9]{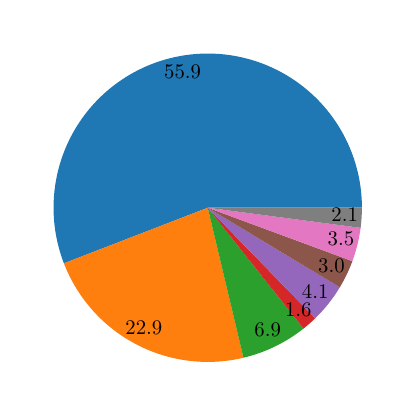}
        \caption{Random dataset.}
        \label{fig:failure-modes-random}
    \end{subfigure}%
    \hfill
    \includegraphics[scale=0.85]{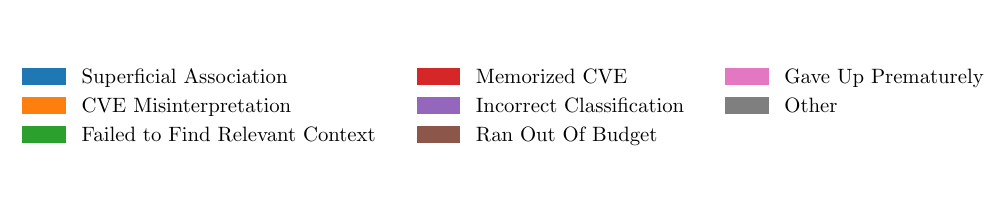}
    \caption{Failure mode reasons.}
        \label{fig:failure-modes}
\end{figure*}

The results are presented in \cref{fig:failure-modes}, showing the dominant source of error is Superficial Association, accounting for 58.7\% of all failures on the realistic dataset and 55.9\% on the random dataset, where the agent infers a valid patch based on surface-level cues such as keyword overlap, file names, or coarse semantic similarity without establishing a causal link between the code changes and the CVE. The second most common failure mode is CVE Misinterpretation, with 29.8\% on realistic dataset, and 22.9\% on the random dataset. This occurs when the agent retrieves the CVE report but misunderstands its root cause, affected component, or exploit mechanism, leading to justifications that do not align with the actual vulnerability description. Together, these two categories account for the vast majority of failures, indicating that incorrect decisions are primarily driven by insufficient semantic grounding rather than lack of exploration or premature termination.

\subsubsection{Stable analysis approach}
\label{sec:rq1-results-stable-analysis-approach}
Comparing agent trajectories across the random and realistic datasets, we observe similar consistent behavioral patterns. This indicates that Favia’s agent exhibits stable and reproducible reasoning behavior, largely unaffected by differences in data distribution and driven primarily by the underlying CVE characteristics.

Comparing the results between \cref{fig:tool-calls-realistic} and \cref{fig:tool-calls-random}, we observe similar agent trajectories and tool usage patterns across the two settings. This indicates that the agent’s exploration strategy is insensitive to the underlying dataset distribution and is instead driven by the CVE description and a fixed reasoning stereotype. Consequently, there no significant performance differences in search depth or tool access. Nevertheless, the variation arises from structural properties of the candidate sets themselves.

In the random dataset, negative samples are typically drawn without semantic or structural proximity to the true patch. As a result, most negative candidates exhibit clear distributional separation from the CVE description, such as unrelated filenames, components, or modification types. This effectively reduces the classification task to a coarse semantic filtering problem, where identifying obvious mismatches is sufficient for correct decisions. Under such conditions, ambiguities or incompleteness in the CVE description are largely masked.

This behavior is illustrated by the random commit from the pytorch-lightning repository associated with CVE-2021-4118, shown in \cref{lst:random-cve-2021-4118}. The CVE description states only that "pytorch-lightning is vulnerable to Deserialization of Untrusted Data." As shown in the listing, the corresponding diff merely corrects a typo in a documentation file. Despite the severe lack of detail in the CVE description, the commit can be readily rejected, as documentation updates are clearly unrelated to a deserialization vulnerability.

\begin{lstlisting}[
    language=diff,
    style=diff,
    caption={Commit 3b6b6c8 randomly selected for CVE-2021-4118.},
    label=lst:random-cve-2021-4118,
    numbers=none,
    frame=none,
    breaklines=true,
    belowskip=0pt,
    escapechar=§]
diff --git a/docs/source/new-project.rst b/docs/source/new-project.rst
index e5ba47351..e50d17046 100644
--- a/docs/source/new-project.rst
+++ b/docs/source/new-project.rst
@@ -339,7 +339,7 @@ You can also add a forward method to do predictions however you want.
\end{lstlisting}
\begin{lstlisting}[
    language=diff,
    style=diff,
    aboveskip=0pt,
    frame=none,
    breaklines=true,
    firstnumber=339,
    escapechar=§]
   {
               return image
   
       autoencoder = LitAutoencoder()
 -     image_sample = autoencoder(§\lstbg{red!40}{(}§)
 +     image_sample = autoencoder()
   
   Option 3: Production
   --------------------
\end{lstlisting}

In contrast, the PatchFinder dataset deliberately constructs hard negatives that are semantically plausible, whose candidates often share keywords, file paths, or conceptual similarity with the CVE description. This collapses the decision boundary from surface-level semantic cues to deep reasoning, requiring the agent to accurately interpret the CVE’s root cause, affected component, and exploit mechanism, and to align them with the specific code changes in the commit. Our failure analysis in Figure \cref{fig:failure-modes} shows that this shift exposes two dominant weaknesses. First, CVE Misinterpretation becomes significantly more frequent, reflecting the difficulty for LLM-based agents to reliably ground under-specified or ambiguous CVE descriptions in code-level semantics. Second, Superficial Association failures increase, indicating that when causal grounding fails, the agent tends to revert to heuristic shortcuts based on keyword overlap or coarse semantic similarity.

Consequently, these results suggest that random negative construction systematically underestimates task difficulty by allowing models to succeed without resolving the core semantic and causal uncertainties inherent in real-world CVE–patch linking. In contrast, PatchFinder-style candidate sets surface these uncertainties explicitly, leading to lower apparent performance but more faithful evaluation of real-world capability.

\vspace{1em}
\noindent\fbox{%
    \parbox{\columnwidth}{%
        \textbf{Summary for RQ2:}
        Agents fail primarily due to insufficient semantic grounding rather than limited exploration or tool access. Most incorrect classifications arise from Superficial Associations, where agents rely on surface-level cues such as keyword overlap, file names, or coarse semantic similarity without establishing a causal link between the code changes and the vulnerability. A secondary but significant source of error is CVE misinterpretation, in which the agent retrieves the CVE report but misunderstands the affected component, root cause, or exploit mechanism. These failures are often amplified by overconfident early termination, where agents commit to a decision without seeking disconfirming evidence, indicating that confidence calibration and causal reasoning—rather than deeper search—are the primary limitations.
    }%
}

\subsection{Results of RQ3: Efficiency}
\label{sec:efficiency}

\begin{figure}[htbp]
    \centering
    \hfill
    \includegraphics[width=\textwidth]{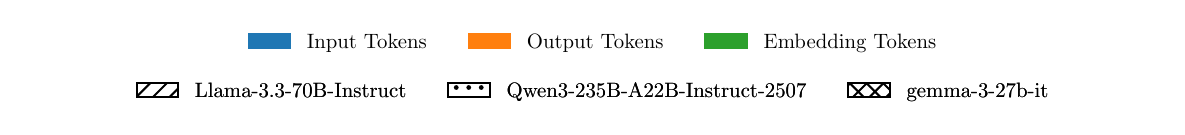}
    \begin{subfigure}[t]{0.5\textwidth}
        \centering
        \includegraphics[scale=0.89]{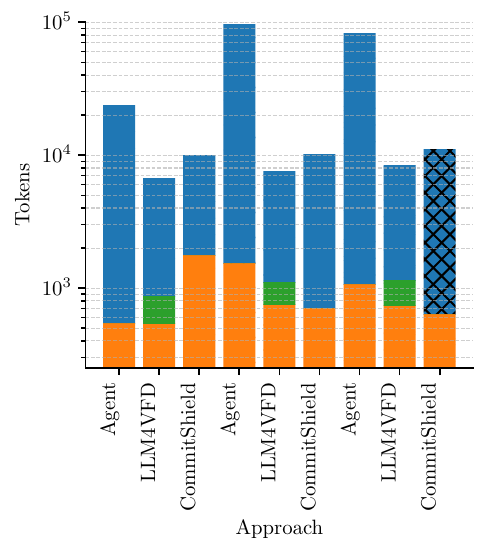}
                \caption{PatchFinder top10 dataset}
        \label{fig:token-usage-random}
    \end{subfigure}%
    \hfill
    \begin{subfigure}[t]{0.5\textwidth}
        \centering
        \includegraphics[scale=0.89]{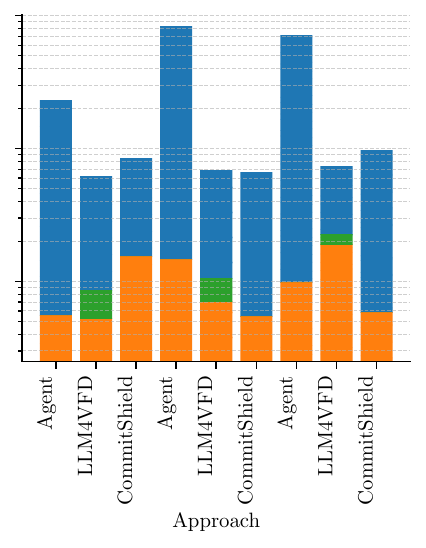}
        \caption{random 10 dataset}
        \label{fig:token-usage-patchfinder}
    \end{subfigure}%
    \caption{Mean token usage for each model using various approaches.}
    \label{fig:token-usage}
\end{figure}

\begin{table}[htbp]
\newcolumntype{Y}{>{\centering\arraybackslash}X}
\newcolumntype{R}{>{\raggedright\arraybackslash}X}
\newcolumntype{L}{>{\raggedleft\arraybackslash}X}
\caption{Mean token usage on realistic dataset across all models.}
\label{tab:token-usage-mean-realistic}
\centering
\begin{tabularx}{\textwidth}{RLLLL}
\hline
\textbf{Method} & \textbf{Input tokens} & \textbf{Output tokens} & \textbf{Embedding tokens} & \textbf{Total tokens} \\
\hline
LLM4VFD & 6,456 & 676 & 362 & 7,494 \\
CommitShield & 9,330 & 1,035 & 0 & 10,365\\
Favia & 66,159 & 1,043 & 0 & 67,202 \\
\hline
\end{tabularx}
\end{table}

\begin{table}[htbp]
\newcolumntype{Y}{>{\centering\arraybackslash}X}
\newcolumntype{R}{>{\raggedright\arraybackslash}X}
\newcolumntype{L}{>{\raggedleft\arraybackslash}X}
\centering
\caption{Token usage on random dataset across all models.}
\label{tab:token-usage-mean-random}
\begin{tabularx}{\textwidth}{RLLLL}
\hline
\textbf{Method} & \textbf{Input tokens} & \textbf{Output tokens} & \textbf{Embedding tokens} & \textbf{Total tokens} \\
\hline
LLM4VFD & 5,368 & 1,040 & 347 & 6,755 \\
CommitShield & 7,328 & 889 & 0 & 8,217\\
Favia & 57,851 & 1,002 & 0 & 58,853 \\
\hline
\end{tabularx}
\end{table}

We measure efficiency using mean token consumption decomposed into input, output, and embedding tokens. The agent operates in a multi-turn fashion so every previous turn and every prior output is appended and counted as new input data; this accumulation is reflected in the input-token totals.

Across the realistic dataset, the agent method Favia consumes substantially more tokens on average than the non-agent baselines. The difference is driven almost entirely by input tokens. As can be seen in \cref{tab:token-usage-mean-realistic}, the mean input tokens for Favia equal 66,159 compared with 6,456 for LLM4VFD and 9,330 for CommitShield. Mean output tokens remain modest and comparable across methods. Favia's mean output tokens equal 1,043 while LLM4VFD and CommitShield report 676 and 1,035 respectively. Embedding tokens are nonzero only for LLM4VFD, which reports a mean embedding cost of 362; embedding overhead is otherwise not applicable.

We see similar usage on the random dataset. From \cref{tab:token-usage-mean-random}, the mean input tokens for Favia, LLM4VFD, and CommitShield is respectively 57,851, 5,368, and 7,328. Favia's mean output tokens equal 1,0402 while LLM4VFD and CommitShield report 1,040 and 889 respectively. Embedding tokens are nonzero only for LLM4VFD, which reports a mean embedding cost of 347.

The elevated input-token cost for the agent follows directly from its multi-turn interaction pattern. Typical agent trajectories begin with retrieval of the CVE description, proceed to file-level localization, and continue with iterative code navigation. Because every prior turn and every prior output is appended and counted as new input, these steps accumulate context across turns and inflate cumulative input tokens relative to single-pass approaches that operate on a fixed shallow context.

Token usage also varies across models. Larger models, such as Qwen3-235B-A22B-Instruct-2507, exhibit the highest token consumption, reflecting longer trajectories and greater context retention across turns. Smaller models, such as gemma-3-27b-it, show lower overall usage while preserving the same qualitative interaction structure. These differences suggest that efficiency is influenced both by the agent framework and by model-specific reasoning behavior.

As shown by the results of RQ1 (see \cref{sec:rq1-results-stable-analysis-approach}), Favia’s reasoning framework is largely insensitive to differences in data distribution and is driven primarily by the characteristics of the underlying CVEs. Consistent with this observation, a comparison of token usage between the realistic and random datasets in \cref{fig:token-usage} reveals the same stable behavior, further confirming that Favia’s token consumption patterns are governed by CVE complexity rather than dataset composition.

Importantly, this increased and stable token consumption is not incidental but necessary to avoid false positives. By enforcing a consistent and CVE grounded evaluation of each candidate commit, independent of superficial security signals or generic robustness improvements, Favia expends additional reasoning effort to verify alignment between the vulnerability description and the affected code. As illustrated in \cref{sec:case-study} from RQ1, this consistency enables Favia to correctly reject commits that appear security relevant at a surface level but are unrelated to the actual vulnerability, in contrast to CommitShield and LLM4VFD. This shows that higher token usage reflects deliberate analytical rigor rather than inefficiency.

Additionally, this disparity in token usage between Favia and the baselines can be translated into real monetary cost under commonly used pay-per-token APIs. As of February 2026, pricing for the latest-generation OpenAI model GPT-5.2 is \$1.75 per million input tokens and \$14.00 per million output tokens, with text-embedding-3-large priced at \$0.13 per million tokens. Using these concrete rates provides a conservative upper-bound estimate, as GPT-5.2 represents one of the most capable and expensive publicly available models. Under this pricing, Favia’s mean usage of 66,159 input tokens and 1,043 output tokens per CVE on the realistic dataset corresponds to an absolute cost of roughly \textbf{\$0.13 per sample}. Non-agent baselines operating in the 7k–10k token range, LLM4VFD would incur approximately \textbf{\$0.02 per sample}, and CommitShield \textbf{\$0.03 per sample}. Although this constitutes a clear relative increase, the absolute difference remains small in practical terms. Because Favia is applied only to a limited set of high-ranked candidate commits (e.g. top-10 per CVE), the total end-to-end cost per vulnerability remains well below one dollar in typical evaluation or deployment scenarios. Consequently, even when instantiated with a state-of-the-art model such as GPT-5.2, the agent’s additional token expenditure is economically feasible and justified by its ability to perform iterative, evidence-driven reasoning and fine-grained code inspection, rather than being a prohibitive barrier at realistic scales.

This trade-off becomes particularly compelling when considering effectiveness. On the realistic dataset, LLM4VFD, CommitShield, and Favia miss a true patch on average (across models) 75, 97, and 33 times, respectively. Thus, Favia identifies on average 42 additional patches compared to the strongest baseline. Since Favia is applied to the top-10 candidate commits per CVE, the cost of fully evaluating one CVE corresponds to ten samples, yielding a per-CVE cost of \(10 \cdot 0.13\$ = 1.30\$\). Recovering these 42 additional patches therefore incurs a total cost of \(42 \cdot 1.30\$ = 54.6\$\). While not negligible, this cost is modest when weighed against the potential consequences of missing a vulnerability-fixing patch, which can entail severe security, financial, and operational risks.

In summary, the agent pays a clear efficiency price in mean input tokens in order to perform iterative, evidence-driven localization and inspection. This additional computational cost supports systematic grounding in CVE descriptions and targeted code inspection rather than increased final verbosity.

\vspace{1em}
\noindent\fbox{%
    \parbox{\columnwidth}{%
        \textbf{Summary for RQ3:} On the realistic dataset the agent Favia incurs a substantially higher mean input-token cost, 66,159 tokens, and a higher mean total, 67,202 tokens, because every prior turn and output is appended and counted as new input. Mean output tokens remain modest at 1,043 and embedding overhead is limited to LLM4VFD at 362. The result is a clear efficiency trade-off: the agent expends more computation to enable iterative, evidence-driven localization while non-agent baselines remain far more token-efficient but operate with shallower, single-pass context.%
    }%
}

\section{Discussion}
\label{chap:discussion}

\subsection{Comparison with Related Studies}
Our findings show that Favia’s agent‑based reasoning provides a substantial advancement over existing vulnerability‑fix detection approaches. Unlike traditional machine‑learning models or embedding‑based LLM systems discussed in \cref{sec:related-work-vfd}, Favia performs iterative, evidence‑grounded reasoning that allows it to correlate CVE semantics with code changes rather than relying on surface‑level similarity. This enables the agent to correctly identify indirect fixes, multi‑file patches, and subtle logic adjustments that prior systems frequently miss. Crucially, Favia can fully automate the final selection of the correct patch, eliminating the manual expert intervention required by ranking‑based systems such as PatchFinder \cite{li2024patchfinder:}, where humans must still inspect the top‑k candidates to determine the true fix.

Our evaluation also highlights a critical gap in recent LLM‑based systems such as LLM4VFD \cite{yang2025code} and CommitShield \citep{wu2025commitshield}. These methods report strong performance but evaluate exclusively on randomly sampled commits, where most negatives are trivially unrelated to security. Such settings dramatically inflate performance and do not reflect real‑world difficulty. By contrast, we evaluate on both heterogeneous random commits (random\_10) and homogeneous, security‑relevant candidate sets (PatchFinder\_top10), where commits are highly similar and difficult to distinguish. This dual evaluation reveals that Favia maintains highest F1 score on all accounts. We also see a significant discrepancy between PatchFinder’s reported recall@10 of 80.63\% and the 48\% recall we observe on our realistic dataset. This further underscores the importance of a sound, rigorous, and representative evaluation dataset which capture the true complexity of vulnerability‑fix identification.

Finally, our efficiency analysis shows that agent‑based reasoning incurs higher computational cost due to multi‑step interactions. However, because Favia evaluates only the top‑k ranked commits, the approach remains practical even for large repositories. Importantly, our candidate‑ranking stage is not tied to any specific method. It can be powered by classical machine‑learning models, lightweight heuristics, or future retrieval systems. Unlike CommitShield \citep{wu2025commitshield}, which relies on language‑specific program‑analysis pipelines and heavyweight static‑analysis infrastructure, Favia performs deep semantic code reasoning while remaining fully language‑agnostic and lightweight. This flexibility allows Favia to operate across diverse repositories without the engineering overhead required by analysis‑dependent systems.

\subsection{Implications to Academia and Industry}
For the research community, our findings highlight two important implications. First, evaluation methodology strongly influences perceived performance. Randomly sampled candidate sets substantially overestimate the effectiveness of all approaches, including LLM-based ones. Realistic, ranked candidate evaluation is therefore essential for meaningful progress in vulnerability-fix detection research. Second, our analysis shows that agent behavior and failure modes can be systematically studied, opening new directions for understanding and improving LLM-based systems beyond aggregate metrics.

For industry practitioners, Favia demonstrates the feasibility of deploying agent-based reasoning in security workflows where precision is critical. While the agent incurs higher computational cost, this cost is concentrated in input tokens used for evidence gathering rather than verbose outputs, aligning well with use cases such as security triage, patch auditing, and dependency risk assessment. Moreover, the agent’s step-by-step reasoning and tool usage provide natural hooks for interpretability and human-in-the-loop validation, which are essential in high-stakes security settings.

\subsection{Threats to Validity}

\subsubsection{External validity}

A potential threat to external validity arises from the selection of models included in our evaluation. We mitigate this risk by evaluating a diverse set of models spanning multiple families, scales, pre-training corpora, and learning objectives. Concerning our experiments, they are conducted on real-world CVEs and corresponding commits drawn from publicly available repositories and spanning multiple programming languages. This substantially improves ecological validity compared to evaluations based on synthetic vulnerabilities or narrowly scoped benchmarks. Because large repositories may contain an extremely high number of commits, processing all commits is computationally and financially infeasible. In line with earlier works \cite{li2024patchfinder:}, we collect a representatively large candidate pool of up to 5,000 commits per repository.

Nevertheless, several limitations remain. First, the construction of realistic candidate sets relies on PatchFinder. While PatchFinder constitutes a strong and widely applicable baseline, alternative retrieval or ranking methods could yield candidate distributions with different characteristics, potentially affecting downstream agent performance. Second, although the dataset spans multiple programming languages, it is skewed toward languages that are well supported by existing tooling and prevalent in CVE reporting, which may limit generalization to less-represented ecosystems. Third, our evaluation is restricted to GitHub-hosted repositories and publicly disclosed CVEs; vulnerabilities in proprietary codebases or those without public advisories may exhibit different patterns that are not captured by our study.

Finally, baseline methods introduce additional external validity considerations. In particular, CommitShield \citep{wu2025commitshield} was originally designed for C and C++ only. Although we extend CommitShield to support C, C++, Python, PHP, Java, and Go, its language coverage remains limited relative to the full diversity of real-world repositories. To ensure a fair and controlled comparison across baselines, we exclude additional auxiliary artifacts such as issue reports and linked discussions.

\subsubsection{Internal validity}

To mitigate bias, we apply consistent experimental protocols across all methods, including identical dataset splits, evaluation metrics, and candidate sets. Dataset splits are performed at the repository level to prevent information leakage between training, validation, and test sets. All baselines are implemented following their original specifications, with deviations explicitly documented when required for scalability or language coverage.

Nonetheless, internal threats remain. First, agent behavior may be sensitive to prompt design, tool availability, and step budgets. While we fix these parameters across experiments, alternative prompts or budgets could influence absolute performance. Second, LLMs may contain latent knowledge of well-known CVEs from pretraining. We partially try to mitigate this risk by analyzing tool usage patterns and identifying memorization-related failure modes, but we cannot fully eliminate the possibility of implicit leakage. We do not add countermeasures for existing approaches, which should mitigate potential bias. Third, failure mode categorization relies on judgments produced by an auxiliary LLM, which may introduce labeling noise despite using a model from an independent family. Here we do however use an model from a completely separate model family in order to produce consistent results.

\section{Conclusion and Future Work}
\label{chap:conclusion}
We introduced Favia, a forensic, agent-based framework for identifying vulnerability-fixing commits that combines scalable candidate ranking with deep, iterative semantic reasoning. Through large-scale evaluation on the CVEVC dataset, we showed that agent-based reasoning consistently outperforms traditional and existing LLM-based approaches under realistic candidate selection, achieving the strongest precision–recall trade-offs and highest F1-scores across models. Our results further demonstrate that evaluations based on random sampling of commits substantially overestimate performance, while realistic candidate sets expose meaningful differences between methods. Most agent failures arise from superficial semantic associations and misinterpretation of CVE descriptions rather than from limited exploration or tool access. Overall, Favia demonstrates that forensic, agent-driven analysis is a practical and effective paradigm for vulnerability-fix detection.

These findings motivate several directions for future work. Improving causal reasoning and confidence calibration may further reduce false positives in ambiguous candidate sets, while adaptive control of reasoning depth could reduce the computational cost of agent-based analysis. More broadly, extending CVEVC and similar datasets to support additional ranking strategies, languages, and vulnerability classes would further enable realistic, scalable evaluation without the prohibitive cost of full repository analysis.

\clearpage

\section*{Acknowledgments}
The empirical results were supported through the computational resources of HPC facilities at the Norwegian University of Science and Technology (NTNU) \citep{sjalander2019epic}.

\section*{Data Availability}
\noindent The cvevc\_candidates dataset is available at:\\
\url{https://huggingface.co/datasets/andstor/cvevc_candidates}\\

\noindent The cvevc\_cve dataset is available at:\\
\url{https://huggingface.co/datasets/andstor/cvevc_cve}\\

\noindent The cvevc\_commit dataset is available at:\\
\url{https://huggingface.co/datasets/andstor/cvevc_commits}\\

\noindent A mapping dataset between the cvevc\_cve and cvevc\_commit dataset is available at:\\
\url{https://huggingface.co/datasets/andstor/cvevc_cve_commit_mappings}\\

\noindent Experiment code and results are available at:\\
\url{https://github.com/andstor/agentic-security-patch-classification-replication-package}\\

\noindent Experiments trajectories are available at:\\
\url{https://huggingface.co/datasets/andstor/favia_trajectories}\\

\noindent Experiments traces are available at following demos:\\
\url{https://huggingface.co/spaces/andstor/phoenix-cvevc_candidates_PatchFinder_top10}\\
\url{https://huggingface.co/spaces/andstor/phoenix-cvevc_candidates_random_10}\\

\printbibliography

@misc{sjalander2019epic,
  title = {{EPIC}: An Energy-Efficient, High-Performance {GPGPU} Computing Research Infrastructure},
  author = {Magnus Sj\"alander and Magnus Jahre and Gunnar Tufte and Nico Reissmann},
  year = 2019,
  eprint = {1912.05848},
  archivePrefix ={arXiv},
  primaryClass = {cs.DC}
}

@inproceedings{tian2012identifying,
	author = {Tian, Yuan and Lawall, Julia and Lo, David},
	booktitle = {2012 34th International Conference on Software Engineering (ICSE)},
	doi = {10.1109/ICSE.2012.6227176},
	issn = {1558-1225},
	month = {6},
	pages = {386-396},
	title = {Identifying Linux bug fixing patches},
	year = {2012}}

@inproceedings{perl2015vccfinder:,
	address = {New York, NY, USA},
	author = {Perl, Henning and Dechand, Sergej and Smith, Matthew and Arp, Daniel and Yamaguchi, Fabian and Rieck, Konrad and Fahl, Sascha and Acar, Yasemin},
	booktitle = {Proceedings of the 22nd ACM SIGSAC Conference on Computer and Communications Security},
	doi = {10.1145/2810103.2813604},
	isbn = {9781450338325},
	location = {Denver, Colorado, USA},
	numpages = {12},
	pages = {426--437},
	publisher = {Association for Computing Machinery},
	series = {CCS '15},
	title = {VCCFinder: Finding Potential Vulnerabilities in Open-Source Projects to Assist Code Audits},
	url = {https://doi.org/10.1145/2810103.2813604},
	year = {2015}}

@inproceedings{zhou2017automated,
	address = {New York, NY, USA},
	author = {Zhou, Yaqin and Sharma, Asankhaya},
	booktitle = {Proceedings of the 2017 11th Joint Meeting on Foundations of Software Engineering},
	doi = {10.1145/3106237.3117771},
	isbn = {9781450351058},
	location = {Paderborn, Germany},
	numpages = {6},
	pages = {914--919},
	publisher = {Association for Computing Machinery},
	series = {ESEC/FSE 2017},
	title = {Automated identification of security issues from commit messages and bug reports},
	url = {https://doi.org/10.1145/3106237.3117771},
	year = {2017}}

@inproceedings{sabetta2018-a-practical,
	address = {Los Alamitos, CA, USA},
	author = {Sabetta, Antonino and Bezzi, Michele},
	booktitle = {2018 IEEE International Conference on Software Maintenance and Evolution (ICSME)},
	doi = {10.1109/ICSME.2018.00058},
	month = sep,
	pages = {579-582},
	publisher = {IEEE Computer Society},
	title = {{ A Practical Approach to the Automatic Classification of Security-Relevant Commits }},
	url = {https://doi.ieeecomputersociety.org/10.1109/ICSME.2018.00058},
	year = {2018}}

@inproceedings{wang2019detecting,
	author = {Wang, Xinda and Sun, Kun and Batcheller, Archer and Jajodia, Sushil},
	booktitle = {2019 49th Annual IEEE/IFIP International Conference on Dependable Systems and Networks (DSN)},
	doi = {10.1109/DSN.2019.00056},
	issn = {1530-0889},
	month = {6},
	pages = {485-492},
	title = {Detecting "0-Day" Vulnerability: An Empirical Study of Secret Security Patch in OSS},
	year = {2019}}

@inproceedings{hoang2019patchnet:,
	author = {Hoang, Thong and Lawall, Julia and Oentaryo, Richard J. and Tian, Yuan and Lo, David},
	booktitle = {Proceedings of the 41st International Conference on Software Engineering: Companion Proceedings},
	doi = {10.1109/ICSE-Companion.2019.00044},
	location = {Montreal, Quebec, Canada},
	numpages = {4},
	pages = {83--86},
	publisher = {IEEE Press},
	series = {ICSE '19},
	title = {PatchNet: a tool for deep patch classification},
	url = {https://doi.org/10.1109/ICSE-Companion.2019.00044},
	year = {2019}}

@inproceedings{machiry2020spider:,
	author = {Machiry, Aravind and Redini, Nilo and Camellini, Eric and Kruegel, Christopher and Vigna, Giovanni},
	booktitle = {2020 IEEE Symposium on Security and Privacy (SP)},
	doi = {10.1109/SP40000.2020.00038},
	pages = {1562-1579},
	title = {SPIDER: Enabling Fast Patch Propagation In Related Software Repositories},
	year = {2020}}

@inproceedings{hoang2020cc2vec:,
	address = {New York, NY, USA},
	author = {Hoang, Thong and Kang, Hong Jin and Lo, David and Lawall, Julia},
	booktitle = {Proceedings of the ACM/IEEE 42nd International Conference on Software Engineering},
	doi = {10.1145/3377811.3380361},
	isbn = {9781450371216},
	location = {Seoul, South Korea},
	numpages = {12},
	pages = {518--529},
	publisher = {Association for Computing Machinery},
	series = {ICSE '20},
	title = {CC2Vec: distributed representations of code changes},
	url = {https://doi.org/10.1145/3377811.3380361},
	year = {2020}}

@article{zhou2021spi:,
	address = {New York, NY, USA},
	articleno = {13},
	author = {Zhou, Yaqin and Siow, Jing Kai and Wang, Chenyu and Liu, Shangqing and Liu, Yang},
	doi = {10.1145/3468854},
	issn = {1049-331X},
	issue_date = {January 2022},
	journal = {ACM Trans. Softw. Eng. Methodol.},
	month = sep,
	number = {1},
	numpages = {27},
	publisher = {Association for Computing Machinery},
	title = {SPI: Automated Identification of Security Patches via Commits},
	url = {https://doi.org/10.1145/3468854},
	volume = {31},
	year = {2021}}

@article{cabrera-lozoya2021commit2vec:,
	author = {Cabrera Lozoya, Roc{\'\i}o and Baumann, Arnaud and Sabetta, Antonino and Bezzi, Michele},
	doi = {10.1007/s42979-021-00566-z},
	id = {Cabrera Lozoya2021},
	issn = {2661-8907},
	journal = {SN Computer Science},
	number = {3},
	pages = {150},
	title = {Commit2Vec: Learning Distributed Representations of Code Changes},
	url = {https://doi.org/10.1007/s42979-021-00566-z},
	volume = {2},
	year = {2021}}

@inproceedings{zhou2021finding,
	author = {Zhou, Jiayuan and Pacheco, Michael and Wan, Zhiyuan and Xia, Xin and Lo, David and Wang, Yuan and Hassan, Ahmed E.},
	booktitle = {2021 36th IEEE/ACM International Conference on Automated Software Engineering (ASE)},
	doi = {10.1109/ASE51524.2021.9678720},
	issn = {2643-1572},
	month = {11},
	pages = {705-716},
	title = {Finding A Needle in a Haystack: Automated Mining of Silent Vulnerability Fixes},
	year = {2021}}

@inproceedings{wang2021patchrnn:,
	author = {Wang, Xinda and Wang, Shu and Feng, Pengbin and Sun, Kun and Jajodia, Sushil and Benchaaboun, Sanae and Geck, Frank},
	booktitle = {MILCOM 2021 - 2021 IEEE Military Communications Conference (MILCOM)},
	doi = {10.1109/MILCOM52596.2021.9652940},
	location = {San Diego, CA, USA},
	numpages = {6},
	pages = {595--600},
	publisher = {IEEE Press},
	title = {PatchRNN: A Deep Learning-Based System for Security Patch Identification},
	url = {https://doi.org/10.1109/MILCOM52596.2021.9652940},
	year = {2021}}

@inproceedings{tan2021locating,
	address = {New York, NY, USA},
	author = {Tan, Xin and Zhang, Yuan and Mi, Chenyuan and Cao, Jiajun and Sun, Kun and Lin, Yifan and Yang, Min},
	booktitle = {Proceedings of the 2021 ACM SIGSAC Conference on Computer and Communications Security},
	doi = {10.1145/3460120.3484593},
	isbn = {9781450384544},
	location = {Virtual Event, Republic of Korea},
	numpages = {18},
	pages = {3282--3299},
	publisher = {Association for Computing Machinery},
	series = {CCS '21},
	title = {Locating the Security Patches for Disclosed OSS Vulnerabilities with Vulnerability-Commit Correlation Ranking},
	url = {https://doi.org/10.1145/3460120.3484593},
	year = {2021}}

@inproceedings{nguyen-truong2022hermes:,
	author = {Nguyen-Truong, Giang and Kang, Hong Jin and Lo, David and Sharma, Abhishek and Santosa, Andrew E. and Sharma, Asankhaya and Ang, Ming Yi},
	booktitle = {2022 IEEE International Conference on Software Analysis, Evolution and Reengineering (SANER)},
	doi = {10.1109/SANER53432.2022.00018},
	pages = {51-62},
	title = {HERMES: Using Commit-Issue Linking to Detect Vulnerability-Fixing Commits},
	year = {2022}}

@article{wu2022-enhancing,
	address = {Los Alamitos, CA, USA},
	author = {Wu, Bozhi and Liu, Shangqing and Feng, Ruitao and Xie, Xiaofei and Siow, Jingkai and Lin, Shang-Wei},
	doi = {10.1109/TDSC.2022.3192631},
	issn = {1941-0018},
	journal = {IEEE Transactions on Dependable and Secure Computing},
	month = jul,
	number = {01},
	pages = {1-15},
	publisher = {IEEE Computer Society},
	title = {{ Enhancing Security Patch Identification by Capturing Structures in Commits }},
	url = {https://doi.ieeecomputersociety.org/10.1109/TDSC.2022.3192631},
	year = {2022}}

@inproceedings{wang2022vcmatch:,
	author = {Wang, Shichao and Zhang, Yun and Bao, Liagfeng and Xia, Xin and Wu, Minghui},
	booktitle = {2022 IEEE International Conference on Software Analysis, Evolution and Reengineering (SANER)},
	doi = {10.1109/SANER53432.2022.00076},
	issn = {1534-5351},
	month = {3},
	pages = {589-600},
	title = {VCMatch: A Ranking-based Approach for Automatic Security Patches Localization for OSS Vulnerabilities},
	year = {2022}}

@inproceedings{nguyen2022vulcurator:,
	address = {New York, NY, USA},
	author = {Nguyen, Truong Giang and Le-Cong, Thanh and Kang, Hong Jin and Le, Xuan-Bach D. and Lo, David},
	booktitle = {Proceedings of the 30th ACM Joint European Software Engineering Conference and Symposium on the Foundations of Software Engineering},
	doi = {10.1145/3540250.3558936},
	isbn = {9781450394130},
	location = {Singapore, Singapore},
	numpages = {5},
	pages = {1726--1730},
	publisher = {Association for Computing Machinery},
	series = {ESEC/FSE 2022},
	title = {VulCurator: a vulnerability-fixing commit detector},
	url = {https://doi.org/10.1145/3540250.3558936},
	year = {2022}}

@article{nguyen2023multi-granularity,
	author = {Nguyen, Truong Giang and Le-Cong, Thanh and Kang, Hong Jin and Widyasari, Ratnadira and Yang, Chengran and Zhao, Zhipeng and Xu, Bowen and Zhou, Jiayuan and Xia, Xin and Hassan, Ahmed E. and Le, Xuan-Bach D. and Lo, David},
	doi = {10.1109/TSE.2023.3281275},
	issn = {1939-3520},
	journal = {IEEE Transactions on Software Engineering},
	month = {8},
	number = {8},
	pages = {4035-4057},
	title = {Multi-Granularity Detector for Vulnerability Fixes},
	volume = {49},
	year = {2023}}

@inproceedings{nguyen2023vffinder:,
	author = {Nguyen, Son and Vu, Thanh Trong and Vo, Hieu Dinh},
	booktitle = {2023 15th International Conference on Knowledge and Systems Engineering (KSE)},
	doi = {10.1109/KSE59128.2023.10299438},
	pages = {1-6},
	title = {VFFINDER: A Graph-Based Approach for Automated Silent Vulnerability-Fix Identification},
	year = {2023}}

@inproceedings{sun2023silent,
	author = {Sun, Jiamou and Xing, Zhenchang and Lu, Qinghua and Xu, Xiwei and Zhu, Liming and Hoang, Thong and Zhao, Dehai},
	booktitle = {Proceedings of the 45th International Conference on Software Engineering},
	doi = {10.1109/ICSE48619.2023.00089},
	isbn = {9781665457019},
	location = {Melbourne, Victoria, Australia},
	numpages = {13},
	pages = {970--982},
	publisher = {IEEE Press},
	series = {ICSE '23},
	title = {Silent Vulnerable Dependency Alert Prediction with Vulnerability Key Aspect Explanation},
	url = {https://doi.org/10.1109/ICSE48619.2023.00089},
	year = {2023}}

@inproceedings{wang2023graphspd:,
	author = {Wang, Shu and Wang, Xinda and Sun, Kun and Jajodia, Sushil and Wang, Haining and Li, Qi},
	booktitle = {2023 IEEE Symposium on Security and Privacy (SP)},
	doi = {10.1109/SP46215.2023.10179479},
	pages = {2409-2426},
	title = {GraphSPD: Graph-Based Security Patch Detection with Enriched Code Semantics},
	year = {2023}}

@inproceedings{li2024patchfinder:,
	address = {New York, NY, USA},
	author = {Li, Kaixuan and Zhang, Jian and Chen, Sen and Liu, Han and Liu, Yang and Chen, Yixiang},
	booktitle = {Proceedings of the 33rd ACM SIGSOFT International Symposium on Software Testing and Analysis},
	doi = {10.1145/3650212.3680305},
	isbn = {9798400706127},
	location = {Vienna, Austria},
	numpages = {13},
	pages = {590--602},
	publisher = {Association for Computing Machinery},
	series = {ISSTA 2024},
	title = {PatchFinder: A Two-Phase Approach to Security Patch Tracing for Disclosed Vulnerabilities in Open-Source Software},
	url = {https://doi.org/10.1145/3650212.3680305},
	year = {2024}}

@misc{chen2024compvpd:,
	archiveprefix = {arXiv},
	author = {Tianyu Chen and Lin Li and Taotao Qian and Jingyi Liu and Wei Yang and Ding Li and Guangtai Liang and Qianxiang Wang and Tao Xie},
	eprint = {2310.02530},
	primaryclass = {cs.CR},
	title = {CompVPD: Iteratively Identifying Vulnerability Patches Based on Human Validation Results with a Precise Context},
	url = {https://arxiv.org/abs/2310.02530},
	year = {2024}}

@inproceedings{wu2025commitshield,
	author = {Wu, Zhaonan and Zhao, Yanjie and Wei, Chen and Wan, Zirui and Liu, Yue and Wang, Haoyu},
	booktitle = {2025 IEEE/ACM 47th International Conference on Software Engineering: Companion Proceedings (ICSE-Companion)},
	doi = {10.1109/ICSE-Companion66252.2025.00082},
	month = apr,
	pages = {279--290},
	publisher = {IEEE},
	title = {COmmitSHield: Tracking Vulnerability Introduction and Fix in Version Control Systems},
	url = {https://doi.org/10.1109/ICSE-Companion66252.2025.00082},
	year = {2025}}

@misc{yang2025code,
	archiveprefix = {arXiv},
	author = {Xu Yang and Wenhan Zhu and Michael Pacheco and Jiayuan Zhou and Shaowei Wang and Xing Hu and Kui Liu},
	eprint = {2501.14983},
	primaryclass = {cs.SE},
	title = {Code Change Intention, Development Artifact and History Vulnerability: Putting Them Together for Vulnerability Fix Detection by LLM},
	url = {https://arxiv.org/abs/2501.14983},
	year = {2025}}

@misc{zhang2025qwen3embeddingadvancingtext,
      title={Qwen3 Embedding: Advancing Text Embedding and Reranking Through Foundation Models}, 
      author={Yanzhao Zhang and Mingxin Li and Dingkun Long and Xin Zhang and Huan Lin and Baosong Yang and Pengjun Xie and An Yang and Dayiheng Liu and Junyang Lin and Fei Huang and Jingren Zhou},
      year={2025},
      eprint={2506.05176},
      archivePrefix={arXiv},
      primaryClass={cs.CL},
      url={https://arxiv.org/abs/2506.05176}, 
}

@inproceedings{sun2024where,
	author = {Sun, Jiamou and Chen, Jieshan and Xing, Zhenchang and Lu, Qinghua and Xu, Xiwei and Zhu, Liming},
	booktitle = {Proceedings of the IEEE/ACM 46th International Conference on Software Engineering},
	collection = {ICSE '24},
	doi = {10.1145/3597503.3639202},
	month = apr,
	pages = {1--13},
	publisher = {ACM},
	series = {ICSE '24},
	title = {Where is it? Tracing the Vulnerability-relevant Files from Vulnerability Reports},
	url = {http://dx.doi.org/10.1145/3597503.3639202},
	year = {2024}}

@misc{yao2023reactsynergizingreasoningacting,
      title={ReAct: Synergizing Reasoning and Acting in Language Models}, 
      author={Shunyu Yao and Jeffrey Zhao and Dian Yu and Nan Du and Izhak Shafran and Karthik Narasimhan and Yuan Cao},
      year={2023},
      eprint={2210.03629},
      archivePrefix={arXiv},
      primaryClass={cs.CL},
      url={https://arxiv.org/abs/2210.03629}, 
}

@inproceedings{yang2024swe-agent,
author = {Yang, John and Jimenez, Carlos E. and Wettig, Alexander and Lieret, Kilian and Yao, Shunyu and Narasimhan, Karthik and Press, Ofir},
title = {SWE-agent: agent-computer interfaces enable automated software engineering},
year = {2024},
isbn = {9798331314385},
publisher = {Curran Associates Inc.},
address = {Red Hook, NY, USA},
booktitle = {Proceedings of the 38th International Conference on Neural Information Processing Systems},
articleno = {1601},
numpages = {125},
location = {Vancouver, BC, Canada},
series = {NIPS '24}
}

@online{linux_cve_announce,
  author = {{Linux Kernel Organization}},
  title = {Linux CVE Announce Archive},
  year = 2025,
  url = {https://lore.kernel.org/linux-cve-announce/2025123055-directory-hemlock-a282@gregkh/},
  urldate = {2025-12-30}
}

@online{llama3_3,
  author = {Meta},
  title = {Llama 3.3 model card},
  year = 2024,
  url = {https://github.com/meta-llama/llama-models/blob/main/models/llama3_3/MODEL_CARD.md},
  urldate = {2026-01-23}
}

@Inbook{Yang2022,
author="Yang, Jeong
and Lee, Young
and McDonald, Arlen P.",
editor="Lee, Roger",
title="SolarWinds Software Supply Chain Security: Better Protection with Enforced Policies and Technologies",
bookTitle="Software Engineering, Artificial Intelligence, Networking and Parallel/Distributed Computing",
year="2022",
publisher="Springer International Publishing",
address="Cham",
pages="43--58",
isbn="978-3-030-92317-4",
doi="10.1007/978-3-030-92317-4_4",
url="https://doi.org/10.1007/978-3-030-92317-4_4"
}

@misc{doll2025unraveling,
      title={Unraveling Log4Shell: Analyzing the Impact and Response to the Log4j Vulnerabil}, 
      author={John Doll and Carson McCarthy and Hannah McDougall and Suman Bhunia},
      year={2025},
      eprint={2501.17760},
      archivePrefix={arXiv},
      primaryClass={cs.CR},
      url={https://arxiv.org/abs/2501.17760}, 
}

@inproceedings{li2017security,
author = {Li, Frank and Paxson, Vern},
title = {A Large-Scale Empirical Study of Security Patches},
year = {2017},
isbn = {9781450349468},
publisher = {Association for Computing Machinery},
address = {New York, NY, USA},
url = {https://doi.org/10.1145/3133956.3134072},
doi = {10.1145/3133956.3134072},
booktitle = {Proceedings of the 2017 ACM SIGSAC Conference on Computer and Communications Security},
pages = {2201–2215},
numpages = {15},
location = {Dallas, Texas, USA},
series = {CCS '17}
}

@article{sliwerski2005when,
	author = {{\'S}liwerski, Jacek and Zimmermann, Thomas and Zeller, Andreas},
	doi = {10.1145/1082983.1083147},
	issn = {0163-5948},
	journal = {ACM SIGSOFT Software Engineering Notes},
	month = may,
	number = {4},
	pages = {1--5},
	publisher = {Association for Computing Machinery (ACM)},
	title = {When do changes induce fixes?},
	url = {http://dx.doi.org/10.1145/1082983.1083147},
	volume = {30},
	year = {2005}}

@inproceedings{kim2008predicting,
	address = {New York, NY, USA},
	author = {Kim, Sunghun and Zimmermann, Thomas and Whitehead, E. James and Zeller, Andreas},
	booktitle = {Proceedings of the 1st India Software Engineering Conference},
	doi = {10.1145/1342211.1342216},
	isbn = {9781595939173},
	location = {Hyderabad, India},
	numpages = {2},
	pages = {15--16},
	publisher = {Association for Computing Machinery},
	series = {ISEC '08},
	title = {Predicting faults from cached history},
	url = {https://doi.org/10.1145/1342211.1342216},
	year = {2008}}

@inproceedings{mockus2000identifying,
	address = {USA},
	author = {Mockus, Audris and Votta, Lawrence G.},
	booktitle = {Proceedings of the International Conference on Software Maintenance (ICSM'00)},
	isbn = {0769507530},
	pages = {120},
	publisher = {IEEE Computer Society},
	series = {ICSM '00},
	title = {Identifying Reasons for Software Changes Using Historic Databases},
	year = {2000}}

@inproceedings{dallmeier2007extraction,
	author = {Dallmeier, Valentin and Zimmermann, Thomas},
	booktitle = {Proceedings of the twenty-second IEEE/ACM international conference on Automated software engineering},
	collection = {ASE07},
	doi = {10.1145/1321631.1321702},
	month = nov,
	pages = {433--436},
	publisher = {ACM},
	series = {ASE07},
	title = {Extraction of bug localization benchmarks from history},
	url = {http://dx.doi.org/10.1145/1321631.1321702},
	year = {2007}}

@article{scikit-learn,
  title={Scikit-learn: Machine Learning in {P}ython},
  author={Pedregosa, F. and Varoquaux, G. and Gramfort, A. and Michel, V.
          and Thirion, B. and Grisel, O. and Blondel, M. and Prettenhofer, P.
          and Weiss, R. and Dubourg, V. and Vanderplas, J. and Passos, A. and
          Cournapeau, D. and Brucher, M. and Perrot, M. and Duchesnay, E.},
  journal={Journal of Machine Learning Research},
  volume={12},
  pages={2825--2830},
  year={2011}
}

@misc{li2022automatingcodereviewactivities,
      title={Automating Code Review Activities by Large-Scale Pre-training}, 
      author={Zhiyu Li and Shuai Lu and Daya Guo and Nan Duan and Shailesh Jannu and Grant Jenks and Deep Majumder and Jared Green and Alexey Svyatkovskiy and Shengyu Fu and Neel Sundaresan},
      year={2022},
      eprint={2203.09095},
      archivePrefix={arXiv},
      primaryClass={cs.SE},
      url={https://arxiv.org/abs/2203.09095}, 
}

@misc{openai2025gptoss120bgptoss20bmodel,
      title={{gpt-oss-120b \& gpt-oss-20b Model Card}}, 
      author={OpenAI and : and Sandhini Agarwal and Lama Ahmad and Jason Ai and Sam Altman and Andy Applebaum and Edwin Arbus and Rahul K. Arora and Yu Bai and Bowen Baker and Haiming Bao and Boaz Barak and Ally Bennett and Tyler Bertao and Nivedita Brett and Eugene Brevdo and Greg Brockman and Sebastien Bubeck and Che Chang and Kai Chen and Mark Chen and Enoch Cheung and Aidan Clark and Dan Cook and Marat Dukhan and Casey Dvorak and Kevin Fives and Vlad Fomenko and Timur Garipov and Kristian Georgiev and Mia Glaese and Tarun Gogineni and Adam Goucher and Lukas Gross and Katia Gil Guzman and John Hallman and Jackie Hehir and Johannes Heidecke and Alec Helyar and Haitang Hu and Romain Huet and Jacob Huh and Saachi Jain and Zach Johnson and Chris Koch and Irina Kofman and Dominik Kundel and Jason Kwon and Volodymyr Kyrylov and Elaine Ya Le and Guillaume Leclerc and James Park Lennon and Scott Lessans and Mario Lezcano-Casado and Yuanzhi Li and Zhuohan Li and Ji Lin and Jordan Liss and Lily and Liu and Jiancheng Liu and Kevin Lu and Chris Lu and Zoran Martinovic and Lindsay McCallum and Josh McGrath and Scott McKinney and Aidan McLaughlin and Song Mei and Steve Mostovoy and Tong Mu and Gideon Myles and Alexander Neitz and Alex Nichol and Jakub Pachocki and Alex Paino and Dana Palmie and Ashley Pantuliano and Giambattista Parascandolo and Jongsoo Park and Leher Pathak and Carolina Paz and Ludovic Peran and Dmitry Pimenov and Michelle Pokrass and Elizabeth Proehl and Huida Qiu and Gaby Raila and Filippo Raso and Hongyu Ren and Kimmy Richardson and David Robinson and Bob Rotsted and Hadi Salman and Suvansh Sanjeev and Max Schwarzer and D. Sculley and Harshit Sikchi and Kendal Simon and Karan Singhal and Yang Song and Dane Stuckey and Zhiqing Sun and Philippe Tillet and Sam Toizer and Foivos Tsimpourlas and Nikhil Vyas and Eric Wallace and Xin Wang and Miles Wang and Olivia Watkins and Kevin Weil and Amy Wendling and Kevin Whinnery and Cedric Whitney and Hannah Wong and Lin Yang and Yu Yang and Michihiro Yasunaga and Kristen Ying and Wojciech Zaremba and Wenting Zhan and Cyril Zhang and Brian Zhang and Eddie Zhang and Shengjia Zhao},
      year={2025},
      eprint={2508.10925},
      archivePrefix={arXiv},
      primaryClass={cs.CL},
      url={https://arxiv.org/abs/2508.10925}, 
}

@misc{gemmateam2025gemma3technicalreport,
      title={Gemma 3 Technical Report}, 
      author={Gemma Team and Aishwarya Kamath and Johan Ferret and Shreya Pathak and Nino Vieillard and Ramona Merhej and Sarah Perrin and Tatiana Matejovicova and Alexandre Ramé and Morgane Rivière and Louis Rouillard and Thomas Mesnard and Geoffrey Cideron and Jean-bastien Grill and Sabela Ramos and Edouard Yvinec and Michelle Casbon and Etienne Pot and Ivo Penchev and Gaël Liu and Francesco Visin and Kathleen Kenealy and Lucas Beyer and Xiaohai Zhai and Anton Tsitsulin and Robert Busa-Fekete and Alex Feng and Noveen Sachdeva and Benjamin Coleman and Yi Gao and Basil Mustafa and Iain Barr and Emilio Parisotto and David Tian and Matan Eyal and Colin Cherry and Jan-Thorsten Peter and Danila Sinopalnikov and Surya Bhupatiraju and Rishabh Agarwal and Mehran Kazemi and Dan Malkin and Ravin Kumar and David Vilar and Idan Brusilovsky and Jiaming Luo and Andreas Steiner and Abe Friesen and Abhanshu Sharma and Abheesht Sharma and Adi Mayrav Gilady and Adrian Goedeckemeyer and Alaa Saade and Alex Feng and Alexander Kolesnikov and Alexei Bendebury and Alvin Abdagic and Amit Vadi and András György and André Susano Pinto and Anil Das and Ankur Bapna and Antoine Miech and Antoine Yang and Antonia Paterson and Ashish Shenoy and Ayan Chakrabarti and Bilal Piot and Bo Wu and Bobak Shahriari and Bryce Petrini and Charlie Chen and Charline Le Lan and Christopher A. Choquette-Choo and CJ Carey and Cormac Brick and Daniel Deutsch and Danielle Eisenbud and Dee Cattle and Derek Cheng and Dimitris Paparas and Divyashree Shivakumar Sreepathihalli and Doug Reid and Dustin Tran and Dustin Zelle and Eric Noland and Erwin Huizenga and Eugene Kharitonov and Frederick Liu and Gagik Amirkhanyan and Glenn Cameron and Hadi Hashemi and Hanna Klimczak-Plucińska and Harman Singh and Harsh Mehta and Harshal Tushar Lehri and Hussein Hazimeh and Ian Ballantyne and Idan Szpektor and Ivan Nardini and Jean Pouget-Abadie and Jetha Chan and Joe Stanton and John Wieting and Jonathan Lai and Jordi Orbay and Joseph Fernandez and Josh Newlan and Ju-yeong Ji and Jyotinder Singh and Kat Black and Kathy Yu and Kevin Hui and Kiran Vodrahalli and Klaus Greff and Linhai Qiu and Marcella Valentine and Marina Coelho and Marvin Ritter and Matt Hoffman and Matthew Watson and Mayank Chaturvedi and Michael Moynihan and Min Ma and Nabila Babar and Natasha Noy and Nathan Byrd and Nick Roy and Nikola Momchev and Nilay Chauhan and Noveen Sachdeva and Oskar Bunyan and Pankil Botarda and Paul Caron and Paul Kishan Rubenstein and Phil Culliton and Philipp Schmid and Pier Giuseppe Sessa and Pingmei Xu and Piotr Stanczyk and Pouya Tafti and Rakesh Shivanna and Renjie Wu and Renke Pan and Reza Rokni and Rob Willoughby and Rohith Vallu and Ryan Mullins and Sammy Jerome and Sara Smoot and Sertan Girgin and Shariq Iqbal and Shashir Reddy and Shruti Sheth and Siim Põder and Sijal Bhatnagar and Sindhu Raghuram Panyam and Sivan Eiger and Susan Zhang and Tianqi Liu and Trevor Yacovone and Tyler Liechty and Uday Kalra and Utku Evci and Vedant Misra and Vincent Roseberry and Vlad Feinberg and Vlad Kolesnikov and Woohyun Han and Woosuk Kwon and Xi Chen and Yinlam Chow and Yuvein Zhu and Zichuan Wei and Zoltan Egyed and Victor Cotruta and Minh Giang and Phoebe Kirk and Anand Rao and Kat Black and Nabila Babar and Jessica Lo and Erica Moreira and Luiz Gustavo Martins and Omar Sanseviero and Lucas Gonzalez and Zach Gleicher and Tris Warkentin and Vahab Mirrokni and Evan Senter and Eli Collins and Joelle Barral and Zoubin Ghahramani and Raia Hadsell and Yossi Matias and D. Sculley and Slav Petrov and Noah Fiedel and Noam Shazeer and Oriol Vinyals and Jeff Dean and Demis Hassabis and Koray Kavukcuoglu and Clement Farabet and Elena Buchatskaya and Jean-Baptiste Alayrac and Rohan Anil and Dmitry and Lepikhin and Sebastian Borgeaud and Olivier Bachem and Armand Joulin and Alek Andreev and Cassidy Hardin and Robert Dadashi and Léonard Hussenot},
      year={2025},
      eprint={2503.19786},
      archivePrefix={arXiv},
      primaryClass={cs.CL},
      url={https://arxiv.org/abs/2503.19786}, 
}

@misc{qwen3technicalreport,
      title={Qwen3 Technical Report}, 
      author={Qwen Team},
      year={2025},
      eprint={2505.09388},
      archivePrefix={arXiv},
      primaryClass={cs.CL},
      url={https://arxiv.org/abs/2505.09388}, 
}

@online{huggingface,
  author = {{Hugging Face}},
  title = {Hugging Face: The AI community building the future},
  year = 2025,
  url = {https://huggingface.co},
  urldate = {2026-02-11}
}

@softwareversion{smolagents:1.21.2,
  title = {`smolagents`: a smol library to build great agentic systems.},
  author = {Aymeric Roucher and Albert Villanova del Moral and Thomas Wolf and Leandro von Werra and Erik Kaunismäki},
  url = {https://github.com/huggingface/smolagents},
  year = {2025},
  version = {1.21.2}
}

@softwareversion{chroma:1.0.20,
  title = {Chroma},
  author = {Chroma},
  url = {https://github.com/chroma-core/chroma},
  year = {2025},
  version = {1.0.20}
}

@softwareversion{joern:1.0.20,
  title = {Joern - The Bug Hunter's Workbench},
  author = {Joern},
  url = {https://joern.io},
  year = {2025},
  version = {v4.0.426}
}

@softwareversion{tree-sitter:0.21.3,
  title = {Tree-sitter},
  author = {Tree-sitter},
  url = {https://tree-sitter.github.io/tree-sitter/},
  year = {2024},
  version = {v0.21.3}
}

\appendix

\section{Prompts Used}
\label{fig:prompt-template-1}

\subsection{System prompt}

\begin{center}
\begin{tcblisting}{
    title=Prompt Template,
    enhanced,
    colbacktitle=black!5!white,
    coltitle=black,
    colback=white,
    colframe=black!75!black,
    boxrule=0.5mm,
    fonttitle=\bfseries,
    fontupper=\ttfamily,
    arc=1mm,
    breakable=true,
    listing only,
    listing options={
        numbers=none,
        frame=none,
        xleftmargin=0pt,
        xrightmargin=0pt,
        columns=flexible,
        breaklines=true,
        breakatwhitespace=false,
        breakautoindent=false,
        breakindent=0pt,
        lineskip=0pt,
        escapechar=§,
        literate={\{}{\{}{1}%
                {\}}{\}}{1}%
      }
}
You are an expert assistant who can solve any task using code blobs. You will be given a task to solve as best you can.
To do so, you have been given access to a list of tools: these tools are basically Python functions which you can call with code.
To solve the task, you must plan forward to proceed in a series of steps, in a cycle of Thought, Code, and Observation sequences.

At each step, in the 'Thought:' sequence, you should first explain your reasoning towards solving the task and the tools that you want to use.
Then in the Code sequence you should write the code in simple Python. The code sequence must be opened with '<code>', and closed with '</code>'.
During each intermediate step, you can use 'print()' to save whatever important information you will then need.
These print outputs will then appear in the 'Observation:' field, which will be available as input for the next step.
In the end you have to return a final answer using the `final_answer` tool.

Here are a few examples using notional tools:
---
Task: "Generate an image of the oldest person in this document."

Thought: I will proceed step by step and use the following tools: `document_qa` to find the oldest person in the document, then `image_generator` to generate an image according to the answer.
<code>
answer = document_qa(document=document, question="Who is the oldest person mentioned?")
print(answer)
</code>
Observation: "The oldest person in the document is John Doe, a 55 year old lumberjack living in Newfoundland."

Thought: I will now generate an image showcasing the oldest person.
<code>
image = image_generator("A portrait of John Doe, a 55-year-old man living in Canada.")
final_answer(image)
</code>

---
Task: "What is the result of the following operation: 5 + 3 + 1294.678?"

Thought: I will use python code to compute the result of the operation and then return the final answer using the `final_answer` tool
<code>
result = 5 + 3 + 1294.678
final_answer(result)
</code>

---
Task:
"Answer the question in the variable `question` about the image stored in the variable `image`. The question is in French.
You have been provided with these additional arguments, that you can access using the keys as variables in your python code:
{'question': 'Quel est l'animal sur l'image?', 'image': 'path/to/image.jpg'}"

Thought: I will use the following tools: `translator` to translate the question into English and then `image_qa` to answer the question on the input image.
<code>
translated_question = translator(question=question, src_lang="French", tgt_lang="English")
print(f"The translated question is {translated_question}.")
answer = image_qa(image=image, question=translated_question)
final_answer(f"The answer is {answer}")
</code>

---
Task:
In a 1979 interview, Stanislaus Ulam discusses with Martin Sherwin about other great physicists of his time, including Oppenheimer.
What does he say was the consequence of Einstein learning too much math on his creativity, in one word?

Thought: I need to find and read the 1979 interview of Stanislaus Ulam with Martin Sherwin.
<code>
pages = web_search(query="1979 interview Stanislaus Ulam Martin Sherwin physicists Einstein")
print(pages)
</code>
Observation:
No result found for query "1979 interview Stanislaus Ulam Martin Sherwin physicists Einstein".

Thought: The query was maybe too restrictive and did not find any results. Let's try again with a broader query.
<code>
pages = web_search(query="1979 interview Stanislaus Ulam")
print(pages)
</code>
Observation:
Found 6 pages:
[Stanislaus Ulam 1979 interview](https://ahf.nuclearmuseum.org/voices/oral-histories/stanislaus-ulams-interview-1979/)
[Ulam discusses Manhattan Project](https://ahf.nuclearmuseum.org/manhattan-project/ulam-manhattan-project/)

(truncated)

Thought: I will read the first 2 pages to know more.
<code>
for url in ["https://ahf.nuclearmuseum.org/voices/oral-histories/stanislaus-ulams-interview-1979/", "https://ahf.nuclearmuseum.org/manhattan-project/ulam-manhattan-project/"]:
    whole_page = visit_webpage(url)
    print(whole_page)
    print("\n" + "="*80 + "\n")  # Print separator between pages
</code>
Observation:
Manhattan Project Locations:
Los Alamos, NM
Stanislaus Ulam was a Polish-American mathematician. He worked on the Manhattan Project at Los Alamos and later helped design the hydrogen bomb. In this interview, he discusses his work at
(truncated)

Thought: I now have the final answer: from the webpages visited, Stanislaus Ulam says of Einstein: "He learned too much mathematics and sort of diminished, it seems to me personally, it seems to me his purely physics creativity." Let's answer in one word.
<code>
final_answer("diminished")
</code>

---
Task: "Which city has the highest population: Guangzhou or Shanghai?"

Thought: I need to get the populations for both cities and compare them: I will use the tool `web_search` to get the population of both cities.
<code>
for city in ["Guangzhou", "Shanghai"]:
    print(f"Population {city}:", web_search(f"{city} population")
</code>
Observation:
Population Guangzhou: ['Guangzhou has a population of 15 million inhabitants as of 2021.']
Population Shanghai: '26 million (2019)'

Thought: Now I know that Shanghai has the highest population.
<code>
final_answer("Shanghai")
</code>

---
Task: "What is the current age of the pope, raised to the power 0.36?"

Thought: I will use the tool `wikipedia_search` to get the age of the pope, and confirm that with a web search.
<code>
pope_age_wiki = wikipedia_search(query="current pope age")
print("Pope age as per wikipedia:", pope_age_wiki)
pope_age_search = web_search(query="current pope age")
print("Pope age as per google search:", pope_age_search)
</code>
Observation:
Pope age: "The pope Francis is currently 88 years old."

Thought: I know that the pope is 88 years old. Let's compute the result using python code.
<code>
pope_current_age = 88 ** 0.36
final_answer(pope_current_age)
</code>

Above example were using notional tools that might not exist for you. On top of performing computations in the Python code snippets that you create, you only have access to these tools, behaving like regular python functions:

<code>
def cve_report(cve_id: string) -> string:
    """This is a tool that fetches information about a CVE (Common Vulnerabilities and Exposures) entry. The information is returned as markdown.

    Args:
        cve_id: The CVE ID to fetch information for.
    """

def cwe_report(cwe_id: string, view: string) -> string:
    """This is a tool that fetches information about a CWE (Common Weakness Enumeration) entry. The information is returned as markdown.

    Args:
        cwe_id: The CWE ID to fetch information for. Example: 79
    """

def file_search(query: string) -> string:
    """This is a tool that searches for files in a Git repository by filename, keyword, or glob pattern. It uses 'git ls-files' to list files tracked by the repository and supports glob-style searches.

    Args:
        query: The filename, keyword, or glob pattern to search for in the repository.
    """

def code_search(query: string, file: string) -> string:
    """Tool for searching for file contents in a git repository files.

    Args:
        query: The string to search for. Can only be simple string.
        file: The file to search in. Optional.
    """

def open_file(path: string) -> string:
    """
        A tool to open a file and initialize it for paginated viewing. The window size is set to 100 lines.
    

    Args:
        path: The path to the file to open.
    """

def scroll_file(direction: string) -> string:
    """
        A tool to scroll up or down through the currently opened file. Scrolls by one window size.
    

    Args:
        direction: The direction to scroll: 'up' or 'down'.
    """

def final_answer(answer: any) -> any:
    """Provides a final answer to the given problem.

    Args:
        answer: The final answer to the problem
    """

</code>

Here are the rules you should always follow to solve your task:
1. Always provide a 'Thought:' sequence, and a '<code>' sequence ending with '</code>', else you will fail.
2. Use only variables that you have defined!
3. Always use the right arguments for the tools. DO NOT pass the arguments as a dict as in 'answer = wikipedia_search({'query': "What is the place where James Bond lives?"})', but use the arguments directly as in 'answer = wikipedia_search(query="What is the place where James Bond lives?")'.
4. Take care to not chain too many sequential tool calls in the same code block, especially when the output format is unpredictable. For instance, a call to wikipedia_search has an unpredictable return format, so do not have another tool call that depends on its output in the same block: rather output results with print() to use them in the next block.
5. Call a tool only when needed, and never re-do a tool call that you previously did with the exact same parameters.
6. Don't name any new variable with the same name as a tool: for instance don't name a variable 'final_answer'.
7. Never create any notional variables in our code, as having these in your logs will derail you from the true variables.
8. You can use imports in your code, but only from the following list of modules: ['collections', 'datetime', 'itertools', 'math', 'queue', 'random', 're', 'stat', 'statistics', 'time', 'unicodedata']
9. The state persists between code executions: so if in one step you've created variables or imported modules, these will all persist.
10. Don't give up! You're in charge of solving the task, not providing directions to solve it.

Now Begin!
\end{tcblisting}

\captionof{figure}{System prompt template.}
\label{fig:system-template}
\bigskip
\end{center}

\subsection{Task prompt}

\begin{center}
\begin{tcblisting}{
    title=Prompt Template,
    enhanced,
    colbacktitle=black!5!white,
    coltitle=black,
    colback=white,
    colframe=black!75!black,
    boxrule=0.5mm,
    fonttitle=\bfseries,
    fontupper=\ttfamily,
    arc=1mm,
    breakable=true,
    listing only,
    listing options={
        numbers=none,
        frame=none,
        xleftmargin=0pt,
        xrightmargin=0pt,
        columns=flexible,
        breaklines=true,
        breakatwhitespace=false,
        breakautoindent=false,
        breakindent=0pt,
        lineskip=0pt,
        escapechar=§,
        literate={\{}{\{}{1}%
                {\}}{\}}{1}%
      }
}
## Task Overview
You are tasked with determining whether the commit `{{ commit_id }}` in the repository `{{ repository }}` is the actual patch for the vulnerability identified as `{{ cve_id }}`.
## Code Changes (from `git show` for `{{ commit_id }}`):
```diff
{{ commit_diff }}
```
You have access to the full codebase at the parent commit of `{{ commit_id }}`. Use all available tools at your disposal to analyze the codebase-such as searching files, reading file contents, navigating function definitions, and locating usage patterns.
---
## 1. Understand the Vulnerability (`{{ cve_id }}`)
Begin by retrieving and then carefully reviewing the official CVE report for `{{ cve_id }}`. The report is your primary source of truth about the nature of the vulnerability. Take the time to read it thoroughly and ensure you understand the context and scope of the issue it describes.
### 1.1 Identify Affected Components
Determine which parts of the system the vulnerability impacts. This may include specific modules, source files, classes, functions, or configuration layers. Understanding what is affected will help guide your investigation into whether the commit addresses these areas.
### 1.2 Understand the Symptoms and Impact
Extract a clear description of what the vulnerability causes when exploited. What is the practical impact? This might include things like privilege escalation, denial of service, authentication bypass, memory corruption, or information disclosure.
### 1.3 Analyze Entry Points
Identify how the vulnerability is triggered. This could be a specific API call, input vector (e.g., HTTP request, user input), or usage pattern. Understanding this helps in locating where and how the system is exposed.
### 1.4 Determine the Root Cause
Look for the underlying flaw in logic, data handling, or assumptions that made the vulnerability possible. Common root causes include missing input validation, unsafe memory operations, race conditions, improper access control, or insecure default configurations.
### 1.5 Optional: Use CWE Classification
If available, use the CVE-to-CWE mapping to categorize the vulnerability into a known class (e.g., buffer overflow, SQL injection, use-after-free). This may give you additional guidance on what kind of fixes to expect in the code.
---
## 2. Analyze the Commit Changes
Next, examine the code changes introduced by the commit `{{commit_id}}`. This will help you form a hypothesis about what the commit is doing and whether it is related to the CVE.
### 2.1 Identify Scope of Changes
Go through the diff and identify what files, functions, methods, classes, constants, or configuration entries have been changed. Pay attention to whether these components overlap with those described in the CVE.
### 2.2 Understand the Purpose of Each Change
Interpret each change carefully. What is the modification doing? Is it fixing a logic bug, adding input validation, tightening access controls, changing types, or restructuring control flow? Make sure you understand the *intended effect* of each code modification.
### 2.3 Map Changes to the CVE Description
Attempt to directly associate each code change with some aspect of the CVE you reviewed in step 1. Does a new check correspond to a missing validation? Does the removal of a function correlate with the described insecure behavior? If a change cannot be tied back to the CVE, be cautious about assuming relevance.
---
## 3. Correlate Code Changes with the CVE
This step is the most critical: you must rigorously validate whether the commit actually addresses and fixes the vulnerability described in the CVE.
### 3.1 Evaluate Direct Fixes
Determine whether the changes mitigate or eliminate the vulnerable behavior. Does the new code prevent the exploit from occurring via the same entry point? Does it neutralize the root cause? Look for structural corrections that align with the vulnerability's nature.
### 3.2 Trace Usage and Context
Investigate how the modified code is used across the repository. Trace function calls, variable usage, and related logic paths. Make sure that the new behavior propagates correctly and consistently throughout the codebase and that it plausibly fixes the described issue.
### 3.3 Confirm Against Symptoms and Entry Points
Verify that the changes prevent the previously identified symptoms from occurring under the same conditions. Use the same entry points and see if the new logic would block or mitigate the attack path.
---
## Common Mistakes to Avoid
Your task is to **verify**, not speculate. Any unjustified assumption is grounds for failure. Do not conclude that a commit is related to a CVE merely because it modifies the same file or function; superficial associations are not sufficient. Similarly, a CVE mention in a commit message does not guarantee that the commit implements a valid fix-only the code changes themselves matter. Do not rely on commit dates to determine relevance, as the actual patch may precede or follow the CVE publication by a significant margin. Finally, be cautious of version bump or release-tag commits: these may indicate when a fix was included but often do not contain the fix themselves. Your analysis should focus strictly on code-level changes that directly and verifiably address the vulnerability.
---
## Final Answer
When you have completed your analysis, please provide your answer in the following format:
<code>
result = {
    "explanation": "Detailed justification including the key evidence you used."
    "confidence": 1 to 5,     # How probable this commit is a patch (see scale below)
    "answer": True or False,  # Is this commit a patch for the CVE?
}
final_answer(result)
</code>
- The answer should be `True` if the commit is a valid patch for the CVE, and `False` otherwise.
- The confidence is an integer score from 1 (lowest) to 5 (highest) based on how strongly the evidence supports your conclusion of the commit being a valid CVE patch. See the confidence scale below for guidance.
- The explanation should clearly explain your reasoning process, mentioning key findings such as how the changes address the CVE's root cause, symptoms, affected components, or entry points.
### Confidence Scale
- 1 = No confidence (little or unclear evidence)
- 2 = Low confidence (some clues, major gaps)
- 3 = Moderate confidence (likely correct, but nontrivial uncertainty)
- 4 = High confidence (strong supporting evidence, small doubts)
- 5 = Full confidence (clear and comprehensive verification)
\end{tcblisting}

\captionof{figure}{Task prompt template.}
\label{fig:task-template}
\bigskip
\end{center}

\end{document}